\documentclass[twocolumn]{aastex631}
\usepackage{graphicx}
\usepackage{epstopdf}
\usepackage{amsmath}
\usepackage{color}
\usepackage{xcolor}
\usepackage{float}
\usepackage{placeins}
\usepackage{booktabs}
\usepackage{newtxtext,newtxmath}
\usepackage{mathtools}
\newenvironment{tightdisplay}
  {\begingroup\small\setlength{\jot}{2pt}\allowdisplaybreaks[1]}
  {\endgroup}
\newcommand\lsim{\mathrel{\rlap{\lower4pt\hbox{\hskip1pt$\sim$}}
\raise1pt\hbox{$<$}}}
\newcommand\gsim{\mathrel{\rlap{\lower4pt\hbox{\hskip1pt$\sim$}}
\raise1pt\hbox{$>$}}}

\shorttitle{Possible Dynamical Pathways to the Misalignment of the VHS 1256-1257 System}
\shortauthors{Holzknecht, Naoz, \& Shariat}

\begin{document}

\title{Possible Dynamical Pathways to the Misalignment of the VHS 1256-1257 System}

\author{Liz Holzknecht}
\affiliation{Department of Physics and Astronomy, University of California, Los Angeles, CA 90095, USA}
\affiliation{Mani L. Bhaumik Institute for Theoretical Physics, Department of Physics and Astronomy, UCLA, Los Angeles, CA 90095, USA}

\author{Smadar Naoz}
\affiliation{Department of Physics and Astronomy, University of California, Los Angeles, CA 90095, USA}
\affiliation{Mani L. Bhaumik Institute for Theoretical Physics, Department of Physics and Astronomy, UCLA, Los Angeles, CA 90095, USA}

\author{Cheyanne Shariat}
\affiliation{Department of Astronomy, California Institute of Technology, 1200 East California Boulevard, Pasadena, CA 91125, USA}

\begin{abstract}
Circumbinary planets (CBPs) provide a unique window into planet formation and dynamical evolution in complex gravitational environments. Their orbits are shaped not only by the protoplanetary disk but also by the perturbations from two stellar hosts, making them sensitive probes of both early- and late-stage dynamical processes. In this work, we investigate the unusual architecture of the VHS J125601.92-125723.9 system, where a retrograde, nearly polar tertiary orbits an extremely low-mass substellar binary in a hierarchical triple configuration. We find that triple body dynamics can naturally reproduce the observed high eccentricity of the inner binary and the tertiary’s near-polar obliquity. However, this configuration alone cannot account for the observed mutual inclination, which is both near-polar and retrograde. This tension suggests two possible formation pathways: either the planet formed in an aligned, protoplanetary disk-like configuration and was later tilted by an additional, undetected fourth companion (below current {\it Gaia} limits), or the system formed close to its current state. Stellar flybys, in contrast, are unlikely due to their long timescales. Our results highlight both the explanatory power and the limitations of triple dynamics, and the potential role of hidden companions in shaping extreme planetary architectures.
%We find that a hidden fourth companion below current Gaia detection limits could generate the required inclination, while stellar flybys are unlikely due to their long timescales. This work highlights the potential role of undetected companions in shaping extreme planetary architectures.

%We explore the origins of misalignment in circumbinary planets (CBPs), using the VHS 1256-1257 system as a test case, and propose three possible origins for such a system. The utility of the VHS 1256 system results from the unique architecture of the system with significant misalignment of the different angular momentum vectors, very high orbital eccentricities, and the mass of the tertiary falling on the super Jupiter-brown dwarf border. The tertiary also occupies a near-polar orbit. We consider the triple body dynamics of the system and propose that hierarchical triple evolution can naturally produce the highly eccentric inner binary and the near-polar tertiary obliquity. We then examine the origins of the large mutual inclination; in particular, we explore the possibility that the system started prograde (mutual inclination $<90^{\circ}$) and was "flipped" to retrograde (mutual inclination $>90^{\circ}$) by a hidden, 4th object in the system, leveraging dynamical stability arguments to constrain the parameter space of such a companion. We also present a scenario where an initially-prograde VHS 1256 b experienced a fly-by encounter that induced a kick, leading to the present, retrograde configuration, and estimate the likelihood of such a kick occurring during the system's lifetime.
\end{abstract}


\section{Introduction}\label{sec:intro}

%\citet{Poon+24} said some things.
%Some things were said \citep[e.g.,][]{Poon+24}. 

Circumbinary planets (CBPs, i.e., Tatooine-like systems) had captured the community’s interest even before their discovery by Kepler \citep[e.g.,][]{Doyle+11,Orosz+12N,Welsh+12,Welsh+15,Kostov+13,Kostov+14,Schwamb+13}. In fact, candidate exoplanets orbiting binary stars had been identified well before the first Kepler detection (Kepler-16). CBPs are expected to be very common \citep[e.g.,][]{Asensio-Torres+18,Martin18,Kostov20b}, since binary stars and planets are both multitudinous \citep[e.g.,][]{Offner+22}. Anecdotally, the fact that circumbinary planets are often among the first to be identified by a new detection technique (pulsar timing, microlensing, transits) already suggests that they are not rare. Based on the first 16 \citep{Baycroft+25} systems detected by {\tt Kepler}, \cite{Armstrong+14} estimated a {\em minimum\/} CBP occurrence rate between about 10\% and 50\% (in the {\tt Kepler} sample), depending critically on the assumed planetary inclination distribution. 

Misalignment of CBP orbits is expected to be a natural consequence of the formation and evolution of close stellar binaries \citep[e.g.,][]{Munoz+15PNAS,Munoz+15,Hamers+15c,Martin+15,Childs+21Polar,Doolin+11,Georgakarakos+24,Smullen+16,Sutherland+19,Chen+19,Chen_Martin+20,Chen_Martin+21}. Moreover, it was shown recently that highly misaligned systems, where the planetary orbit is nearly perpendicular to the binary orbital plane, can be stable configurations \citep{Naoz+17,Zanardi+17,Vinson+18,DeElia+19}. Circumbinary discs in a polar orientation have been observed, and any planets formed from these discs would be misaligned \citep[][]{Cuello+19, Kennedy+19}. These so-called ``polar orbits'' \citep{Far+10} involve strong secular coupling between the planet and the inner binary. 
%Arguments from conservation of angular momentum and planetary formation indicate that it is unlikely for a planet to form with a retrograde spin-orbit angle with respect to the star (or binary pair of stars) it orbits \citep{Li+16, Foucart+13, Foucart+14, Rosenfeld+13, Czekala+15,Czekala+16, Czekala+19, Kennedy+12b}, but see \citet{Kennedy+12a}. 

%, a prospect that invites additional intrigue given that the system is quite young with an age of $140$ Myr. The high eccentricity of the inner binary is also interesting to note, as it may be indicative of EKL eccentricity spikes.

\begin{figure}[t]
 \centering
 \includegraphics[width=0.5\textwidth]{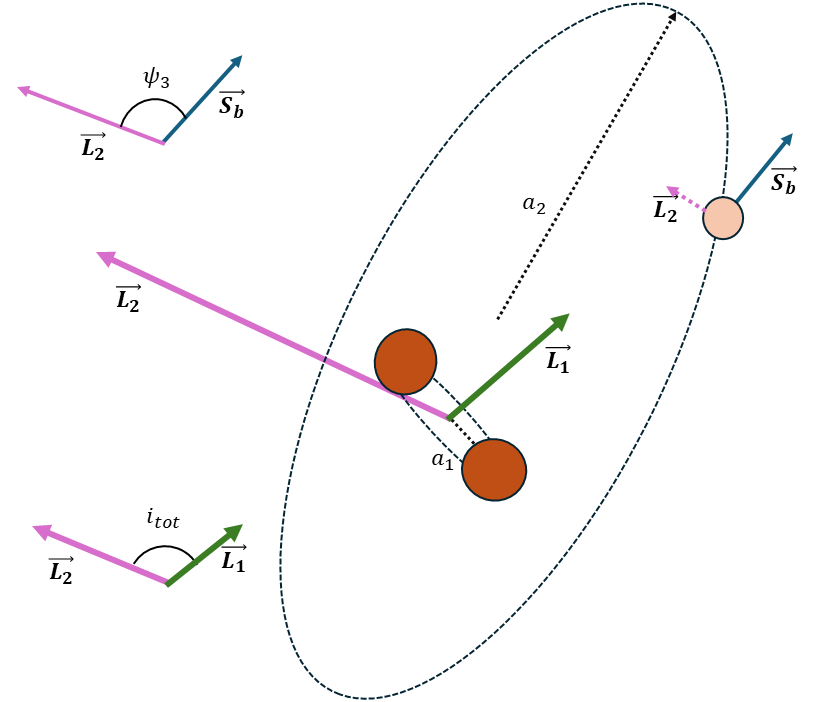}
 \caption{A cartoon (not to scale) of the VHS 1256-1257 system depicting the hierarchical configuration and angles of interest. The frame of reference is the invariable plane. $\vec{L}_1$ ($\vec{L}_2$) is the orbital angular momentum of the inner (outer) orbit; The mutual inclination, $i_{\rm tot}$, is the angle between these two orbital angular momentum vectors. $\vec{S}_b$ is the tertiary spin axis, and $\psi_3$, the tertiary obliquity, is the angle between the planet's spin axis and the outer orbital angular momentum vector. The semimajor axes (SMA) are $a_1$ for the inner binary and $a_2$ for the outer binary. }
 \label{fig:cartoon}
\end{figure}

Against this backdrop, we focus on a unique test system called the VHS J125601.92-125723.9 system (hereafter referred to as VHS 1256). This system is composed of two low-mass (total mass $~0.141 \pm 0.008$~M$_{\odot}$), highly eccentric brown dwarfs orbited by a retrograde, nearly polar massive super-Jupiter \citep{poon+24, Dupuy+23, Stone+16, Gauza+15, Rich+16, Zhou+20, Bowler+20, Miles+23} in a hierarchical triple configuration (see Figure \ref{fig:cartoon} for an illustration). The inner binary is very eccentric ($e_{1}\sim0.883$) with a semimajor axis (SMA) $a_{1}$ of $1.96 \pm 0.03$~au, while the tertiary companion has a SMA $a_{2}$ of $\sim383$~au and eccentricity $e_{2}$ of $\sim0.7$ \citep{Dupuy+23, poon+24}. \citet{poon+24} measured the rotation period of $22.04 \pm 0.05$ hours for VHS 1256 b, the tertiary. Further, \citet{Dupuy+23} presents a bimodal distribution for the mass of VHS 1256 b, peaking at $\sim12$~M$_{\rm J}$ and $\sim16$~M$_{\rm J}$; the bimodality of this distribution arises because the tertiary's mass and age are consistent with evolutionary tracks for both a deuterium inert model and a deuterium fusing model \citep{Dupuy+23}. See  Table \ref{tab:vhsobs} for the full orbital parameters of the system.  

The mutual inclination between the two orbits $i_{\rm tot}$, which is defined as the angle between the orbital angular momentum vectors of the inner and outer binaries, is found to be $ 118^{+12^{\circ}}_{-16^{\circ}}$ \citep[][]{Stone+16, poon+24, Dupuy+23, Zhou+20, Bowler+20, Miles+23}. The VHS 1256 system is unique in that it contains an exoplanet with a measured obliquity. Interestingly, VHS 1256 b's obliquity $\psi_3$ is also measured to be nearly polar, specifically with $90^{\circ}\pm25^{\circ}$, \citep{poon+24}. Thus, the large, retrograde mutual inclination, the near-polar obliquity, and the hierarchical configuration of the system as a whole suggest a rich dynamical history.

%a prospect that invites additional intrigue given that the system is quite young with an age of $140$ Myr. The high eccentricity of the inner binary is also interesting to note, as it may be indicative of EKL eccentricity spikes.

%In Figure \ref{fig:cartoon}, we highlight the hierarchical configuration of the system and denote the angles of interest. The orbital parameters from \cite{poon+24}  for the system are listed in Table \ref{tab:vhsobs}. 

\begin{table}[h!]
\centering
\caption{Parameters for the VHS 1256-1257 system \citep{Dupuy+23, poon+24}}
\label{tab:vhsobs}
\begin{tabular}{ll}
\toprule
\textbf{Parameter} & \textbf{Measured Value} \\
\midrule
VHS 1256 AB mass & $0.141 \pm 0.008$~M$_{\odot}$ \\
VHS 1256 AB SMA & $1.96 \pm 0.03~\mathrm{au}$ \\
VHS 1256 AB eccentricity & $0.883 \pm 0.003$ \\
VHS 1256 b mass & $12.0 \pm 0.1$~M$_{\rm J}$ or $16 \pm 1$~M$_{\rm J}$ \\ \\
VHS 1256 b SMA & $350^{+110}_{-150}~\mathrm{au}$ \\ \\
VHS 1256 b eccentricity & $0.7 \pm 0.1$ \\
VHS 1256 b obliquity & $90^{\circ} \pm 25^{\circ}$ \\ \\
VHS 1256 mutual inclination & $118^{+12^{\circ}}_{-16^{\circ}}$ \\ \\
VHS 1256 Age & $140 \pm 20~\mathrm{Myr}$ \\
\bottomrule
\end{tabular}
\end{table}

The retrograde, nearly polar mutual inclination as well as the nearly polar obliquity of the tertiary invite additional intrigue given that the system is relatively young, with \citet{Dupuy+23} finding a cooling age for the system of $140\pm20$~Myr. Retrograde and high obliquity planets are a natural consequence of high eccentricity migration involving a far-away companion \citep{Naoz+12, Naoz+14, Weldon+25, Weldon+24b, Weldon+24, Weldon+25a}. This process is also known as the Eccentric Kozai-Lidov (EKL) mechanism \citep[][]{Kozai62,Lidov62,Naoz16}.  However, in the VHS 1256 system, the tertiary is less massive than the inner binary, and thus is somewhat inefficient in torquing the inner binary \citep{Dupuy+23, Stone+16, Naoz+11, Naoz+13}. Specifically, \citet{poon+24} recently suggested that dynamical mechanisms such as EKL or a stellar flyby are insufficient to explain both the significant mutual inclination and the large obliquity.  However, we find that EKL can explain the high eccentricity of the inner binary and the near-polar obliquity of VHS 1256 b. While EKL effects can cause the inclination of a VHS 1256-like system to flip from prograde (ie, $i_{\rm tot}<90^{\circ}$) to retrograde (ie, $i_{\rm tot}>90^{\circ}$), we find that these flips are insufficient to explain the $\sim118^{\circ}$ mutual inclination.  Thus, we explore the possibility of a stellar flyby in depth to investigate the origin of the large mutual inclination.

In this work, we utilize VHS 1256 as a case study to understand the role of dynamical interactions in shaping the architecture of systems like this. Particularly, we run a simulation suite containing a large set of VHS 1256-like systems and demonstrate that a near-polar obliquity naturally takes place in such a configuration. On the other hand, while some systems flip their mutual inclination via EKL, it has a very limited range due to the weak torques.  Specifically, any system that evolved to have a mutual inclination consistent with the observations started with an initially retrograde inclination. We thus hypothesize that there are three possibilities for the origin of the observed misalignment: the system formed misaligned via core/filament fragmentation and subsequent gas-driven migration and dynamical evolution \citep{Lee+19, poon+24}, the system formed with the tertiary on a prograde orbit and was excited to a retrograde inclination via dynamical interactions with a hidden fourth companion, or the system formed prograde and the inclination was flipped by an impulse generated from a stellar flyby early in the system's history.

The paper is organized as follows: in Section \ref{sec:triple}, we discuss the triple body dynamics of the system and propose that the high eccentricity and near-polar obliquity of VHS 1256 b can be explained by the triple evolution of the system. In Section \ref{sec:origins}, we explore three possible origins of the retrograde mutual inclination: Section \ref{ssec:4body} considers the possibility that the system formed prograde and was flipped by a hidden fourth companion, Section \ref{ssec:flyby} considers a possible stellar flyby early in the system's lifetime, and Section \ref{ssec:filament} explores filament fragmentation. In Section \ref{sec:discussion}, we summarize our findings and lay the groundwork for future studies of systems containing circumbinary planets.  

%\vspace{0.2cm}

\section{Exploring the triple evolution of the system}\label{sec:triple}

As a first step, we analyze the triple body evolution of the system to constrain the range of initial conditions that yield the observational parameters. We integrate the equations of motion up to the hexadecapole level of approximation \citep{Will+17, Hamers18, Conway+24, Will21}; see Appendix \ref{appendix:hex} for the full set of equations. The utility of expanding beyond the octupole term to include the hexadecapole level is largely twofold. The octupole term is proportional to the mass difference between the members of the inner binary \citep{Naoz16}, rendering it insufficient to analyze systems with a nearly equal mass inner binary such as the VHS 1256 system explored here. Further, the dynamics of a system where the ratio between the outer orbital period and the EKL timescale is comparable or greater than the strength of the octupole term are more accurately captured at the hexadecapole level  \citep[e.g.,][]{Soderhjelm75,Cuk+04,Luo+16,Will+17,Will21,Tremaine23,Klein+24}.
%There are two main reasons for the usefulness of this approximation beyond the octupole level. The first is that it allows us to integrate inner binaries with comparable masses. The second reason relates to the fact that in some of the systems, the ratio between the period of the outer orbit and the timescale of the EKL cycles is comparable to or larger than the strength of the octupole term. It was demonstrated that in such systems, the octupole level is insufficient, and the next-level approximation, i.e., hexadecapole, allows for a more accurate description of the dynamics  \citep[e.g.,][]{Soderhjelm75,Cuk+04,Luo+16,Will17,Will21,Tremaine23,Klein+24}.  

\begin{figure*}
  \begin{center}
    \includegraphics[width=1.0\textwidth]{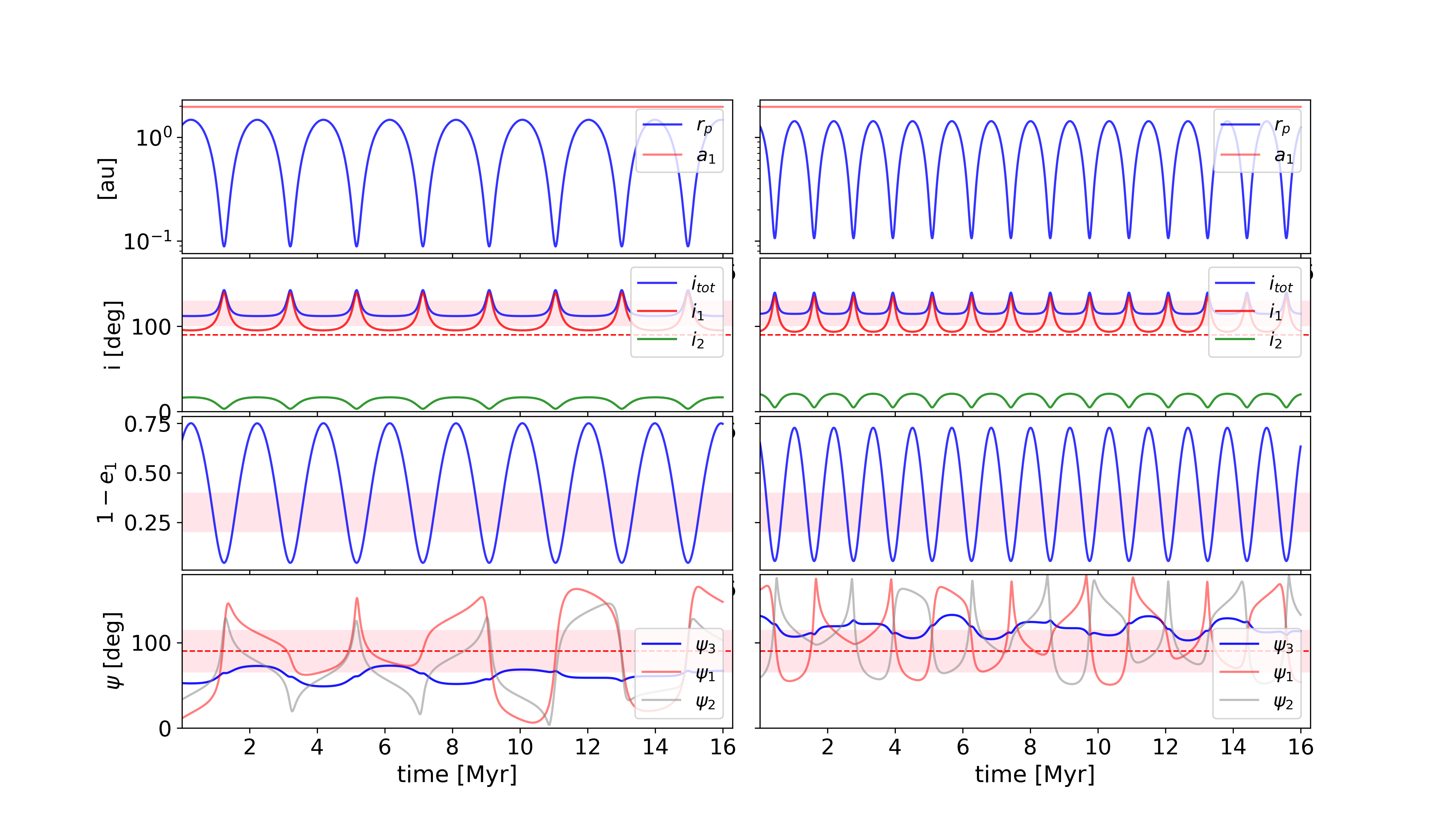}   
    \par\vspace{1ex}
    \parbox{\textwidth}{ % Match figure width
      \caption{\upshape 
      {\bf Left:} An example time evolution that landed in the target region. 
      We note that the eccentricity was excited from an initial value of $e_{1i}=0.33$, demonstrating that the final inner orbit eccentricity is not strongly dependent on the initial value. 
      The inclination remains fairly stable, other than small spikes that coincide with the eccentricity excursions. The initial tertiary obliquity for this system was $\psi_{3i}=52.6^{\circ}$. {\bf Right:}  An example time evolution for one of the samples that landed in the target region. We note that the eccentricity was excited from an initial value of $e_{1i}=0.35$. The inclination was stable except for the peaks corresponding to the eccentricity excursions. In this system, the initial tertiary obliquity was $\psi_{3i}=131.4^{\circ}$. }
      \label{fig:Ex1_orbit}
    }
  \end{center}
\end{figure*}

Additionally, we include the general relativity (GR) precession of the inner and outer orbits \citep[e.g.,][]{Naoz+13b}, equilibrium tidal effects on the inner binary members \citep[e.g.,][]{Eggleton98, Hut1980, Kiseleva+98}, the spin precession of the stars, and notably for the spin evolution of the tertiary (see Appendix \ref{appendix:spin} for the spin equations of the tertiary). The spin precession is incorporated into our simulations through the standard rigid-body torque equations. Due to the size of the outer orbit, the obliquity evolution for the tertiary is dominated by spin precession; the tidal evolution is negligible in this case. 

 We expect that the planet's obliquity will be naturally excited to its observed near-polar configuration for such a system, as depicted in the example system in Figure \ref{fig:Ex1_orbit}. In the system depicted in the left panel, the initial obliquity is $\psi_{3i}=52.6^{\circ}$, reaching a maximum value of $\sim75^{\circ}$, well within the observational range (over-plotted as the pink band). The right panel depicts a system that started with $\psi_{3i}=131.4^{\circ}.$ As can be seen, torquing the mutual inclination is somewhat weak, as expected from the iEKL-like behavior, where iEKL refers to the perturbations induced by the inner binary on the tertiary. However, unlike the full outer orbit test particle case, the planet here is sufficiently massive compared to the inner binary members that it can drive the star's eccentricity to high values, as suggested by \citet{Dupuy+23}. Notably, in both cases, $e_{1}$ was excited from a small initial value ($0.33$ and $0.35$, respectively). The evolution of the inclination demonstrates the characteristic spikes associated with the high eccentricity excursions, but maintains its initial retrograde status.

\begin{figure*}[t]
 \centering
 \includegraphics[width=\textwidth]{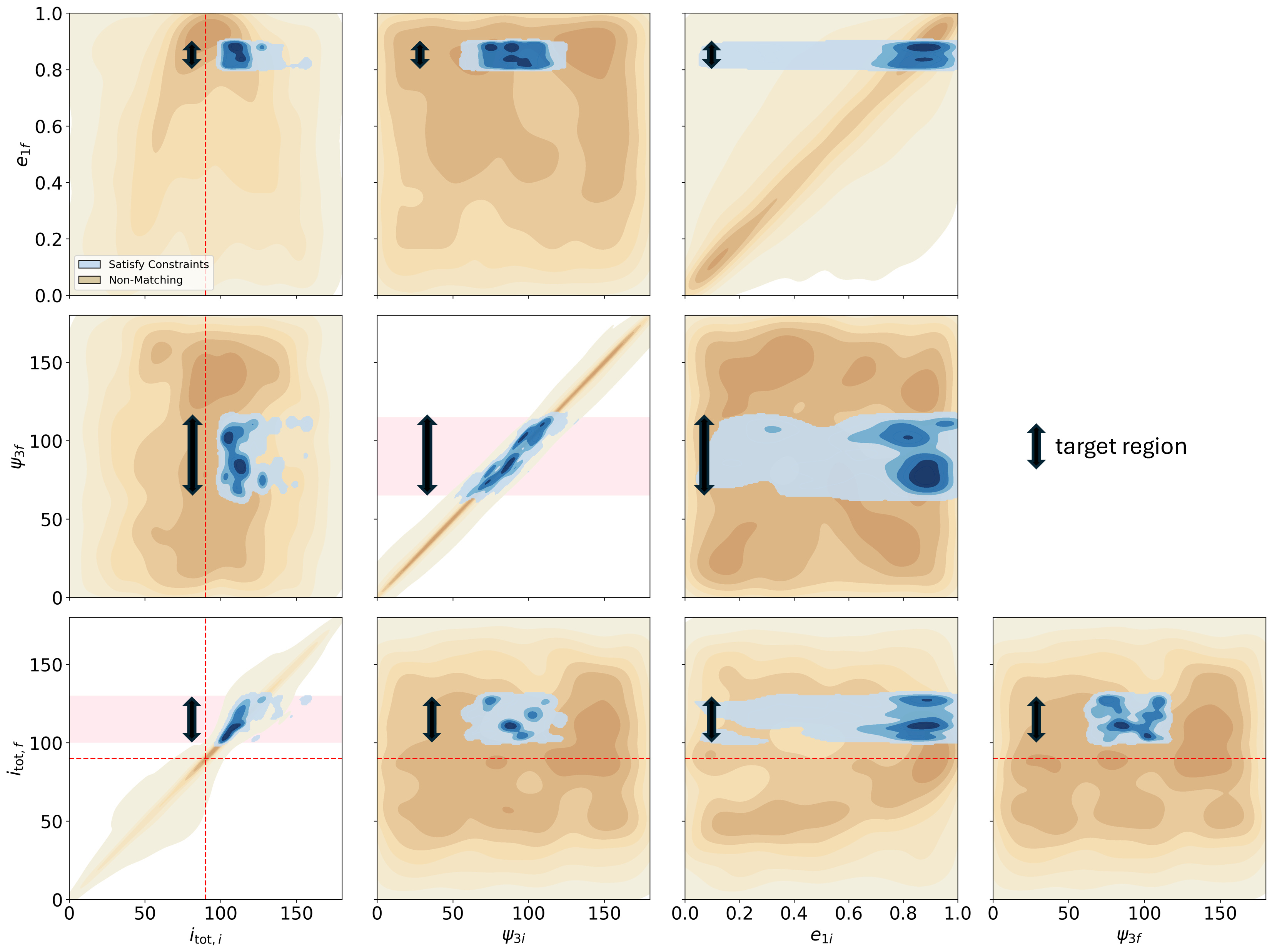}
 \caption{The evolution outcome of the 2000 sample systems with $q=\frac{m_{\rm A}}{m_{\rm B}}\in[0.9,1.0]$ and $m_{3}\in[15$~M$_{\rm J},17$~M$_{\rm J}]$ corresponding to a deuterium inert-model for VHS 1256 b. Here, we show in total $16\times 10^5$ realizations $120-160$~Myrs.  The red dashed lines indicate $90^{\circ}$. The pink shaded boxes highlight the "target region", where we define the target region based on the observational constraints, $e_{1}\in[0.8,0.9]$, $i_{\rm tot}\in[100^{\circ},130^{\circ}]$, and $\psi_{3}\in[65^{\circ},115^{\circ}]$. For each system, $400$ timestamps are considered, ranging from $120$ to $160$~Myr. If at a given timestamp, the system obeys all three observational constraints, the point turns blue. If it does not, it turns beige. Once all timestamps are considered for each system, the blue points are accumulated and displayed by the blue density clouds, while the beige points are accumulated and displayed by the beige density clouds.}
 \label{fig:timestamps16}
\end{figure*}

To test this scenario further, we integrate $4000$ systems for $160$~Myr, where half of these samples have the mass of the tertiary chosen from a uniform distribution between $15-17$~M$_{\rm J}$ while the other half use masses chosen from a uniform distribution between $11-13$~M$_{\rm J}$, corresponding to the two values found by \citet{Dupuy+23}. The results are depicted in Figure \ref{fig:timestamps16} and Figure  \ref{fig:timestamps12} in Appendix \ref{appendix:m312}, respectively.
%We simulate 4000 sample systems and integrate our simulations for 160 Myr. 
Since the age of the system is estimated to be $140\pm20$~Myr, we examine $400$ snapshots between $120$ and $160$~Myr; thus, overall, we have $\sim16\times 10^5$ data points. 
%to determine which systems match the observed tertiary obliquity, inner binary eccentricity ($e_{1}$), and mutual inclination during the lifetime of the system.

The initial conditions for the Monte Carlo simulations are chosen as follows:
The obliquities of all three members of the system ($\psi_{1}, \psi_{2}, \psi_{3}$) are chosen from uniform distribution between $0-\pi$. The arguments of periapsis of the inner and outer orbits are chosen from a uniform distribution between $0-2\pi$. % {\sn{Liz, please continue to do the rest}} 
The spin period of the tertiary is chosen from a normal distribution centered at $22$ hours with a standard deviation of $0.05$ hours \citep{poon+24}, whereas the spin of each brown dwarf in the inner binary is chosen to be $10$ hours \citep{Tannock+21}; we note that the choice of these initial values has negligible effects on the dynamics as the spin mainly affects the tidal evolution, which is inconsequential here.  The total mass of the inner binary is chosen to be $0.141$~M$_{\odot}$; the mass ratio $q$ of the components is chosen from a uniform distribution between $0.9-1$ to reflect the nearly equal apparent magnitudes of the two components measured by \citet{Stone+16}. We choose the eccentricity of the outer orbit from a uniform distribution between $0.6$ and $0.8$, consistent with the measured value in \citet{Dupuy+23} and \citet{poon+24}. We choose the eccentricity of the inner orbit from a uniform distribution between $0$ and $1$, as we note that the triple evolution of the system can excite the inner binary's eccentricity due to the significant mass ($\sim16$~M$_{\rm J}$) of the tertiary \citep{Stone+16, poon+24, Dupuy+23}. 

Our choices for the semimajor axes are motivated by the observations in \citet{Stone+16, Dupuy+23} and \citet{poon+24}: the inner binary SMA is chosen to be $1.96$~au, and the outer orbit SMA is chosen from a uniform distribution between $200-460$~au. Finally, we choose the mutual inclination from a distribution uniform in $\cos(i_{\rm tot})$ between $0-\pi$ to examine the possibility of an initially prograde system flipping to the observed retrograde configuration. We analyze our simulation data using the observational constraints for $i_{\rm tot}$, $e_{1}$, and $\psi_{3}$ to attempt to reproduce the unique relationships between the different angular momentum vectors in this system (i.e., the two orbit normals and the tertiary spin axis). $e_1$ can change dramatically throughout the system's evolution via EKL; thus, the $e_1$ constraint ensures that any candidate systems match the observed orbital architecture of the inner binary \citep{poon+24, Dupuy+23}

% Cheyanne up to here
Figure \ref{fig:timestamps16} presents the results of $16\times 10^5$ outcomes of the dynamical integration of the system. Specifically, we consider the obliquity of the tertiary, the mutual inclination, and the inner binary eccentricity. If at a given timestamp, the system obeys all three observational constraints, the point turns blue. If it does not, it turns beige. Once all timestamps are considered for each system, the blue
points are accumulated and displayed by the blue density clouds, while the beige points are accumulated and displayed by the beige density
clouds. 

As depicted, EKL can naturally drive the eccentricity of the inner binary to the observed value for a wide range of initial conditions. Thus, the configuration of the system is not strongly dependent on the initial eccentricity of the inner binary. This behavior is shown in both mass range simulations (see Appendix \ref{appendix:m312}).  
%We notice this in both sets of simulations corresponding to the different mass ranges. 
Similarly, a fairly wide range in the initial conditions for $\psi_{3}$ dynamically evolved into the final configuration. The target region for  $\psi_{3}$ represents the error bars on the observed measurement. As depicted in Figure 3, several systems that started with an obliquity outside of the target region evolved to the observed value. Specifically, the minimum initial obliquity that evolved into the target region is $\sim 50^{\circ}$, as highlighted by the blue contours in the middle column of Figure 3.

%The error bars on the observed measurement for $\psi_{3}$ are large, so the target region itself is much larger. Nonetheless, several systems that started with an obliquity outside of the target region evolved to the observed value.  Specifically, while some systems with an initially prograde obliquity less than $90^{\circ}$ evolved to match the observed range for $\psi_3$, we note that in our simulations, the minimum initial obliquity that evolved into the target region was approximately $50^{\circ}$, as highlighted by the blue contours in the middle column of Figure 3. 

While the obliquity of smaller tertiary planets can be excited from near-zero to near-polar via triple-body dynamics, the tertiary's significant mass in the VHS 1256 system limits the parameter space for this flip. Thus, we find a range of initial obliquities from $\sim50^{\circ}$ to $\sim130^{\circ}$ that were able to evolve to match the observed configuration. Non-zero obliquities are observed even within our own solar system; therefore, this minimum obliquity is not a prohibitive constraint on the initial conditions for the tertiary. Thus, the highly eccentric inner binary and the tertiary obliquity can be understood as dynamical consequences of hierarchical triple evolution.

The mutual inclination is strongly dependent on the initial value, as depicted in the left-most panels of Figure  \ref{fig:timestamps16}. One possible dynamical origin discussed and ultimately refuted via a timescale argument in \citet{poon+24} is the idea that VHS 1256 b formed closer to the binary in a prograde configuration but was ``kicked out'' onto an inclined orbit with a large obliquity via EKL. Consistently, we find that all of the systems that match the observed values for the mutual inclination, obliquity, and eccentricity started with a mutual inclination above $90^{\circ}$; in other words, it is unlikely that the orbital flip was generated by triple body interactions between the tertiary and VHS 1256AB, which we understand as a consequence of the mass of the tertiary. Thus, either VHS 1256 b formed with its current inclination via filament fragmentation that effectively randomized the angular momentum vectors \citep[e.g.,][]{poon+24}, or it formed in a prograde configuration and was flipped by an interaction with another massive body. Although triple dynamics are capable of generating some of the observed parameters of this system (i.e., the tertiary obliquity and high eccentricity of the inner binary), they cannot reproduce all properties of this system. For this particular system, the significant mass of the tertiary limits the ability of the inner binary to torque the outer orbit and excite the mutual inclination. The large semimajor axis of the outer orbit produces further constraints, as it limits the range of initial tertiary obliquities that can evolve into the observed regime. With these limitations in mind, the analysis of the VHS 1256-1257 system provides a unique case study into the triple dynamics that operate in systems with massive tertiary companions on very wide, eccentric orbits. We discuss the possible origin of the system in Section \ref{sec:origins}.

\begin{figure*}
 \begin{center}
 \includegraphics[width=\textwidth]{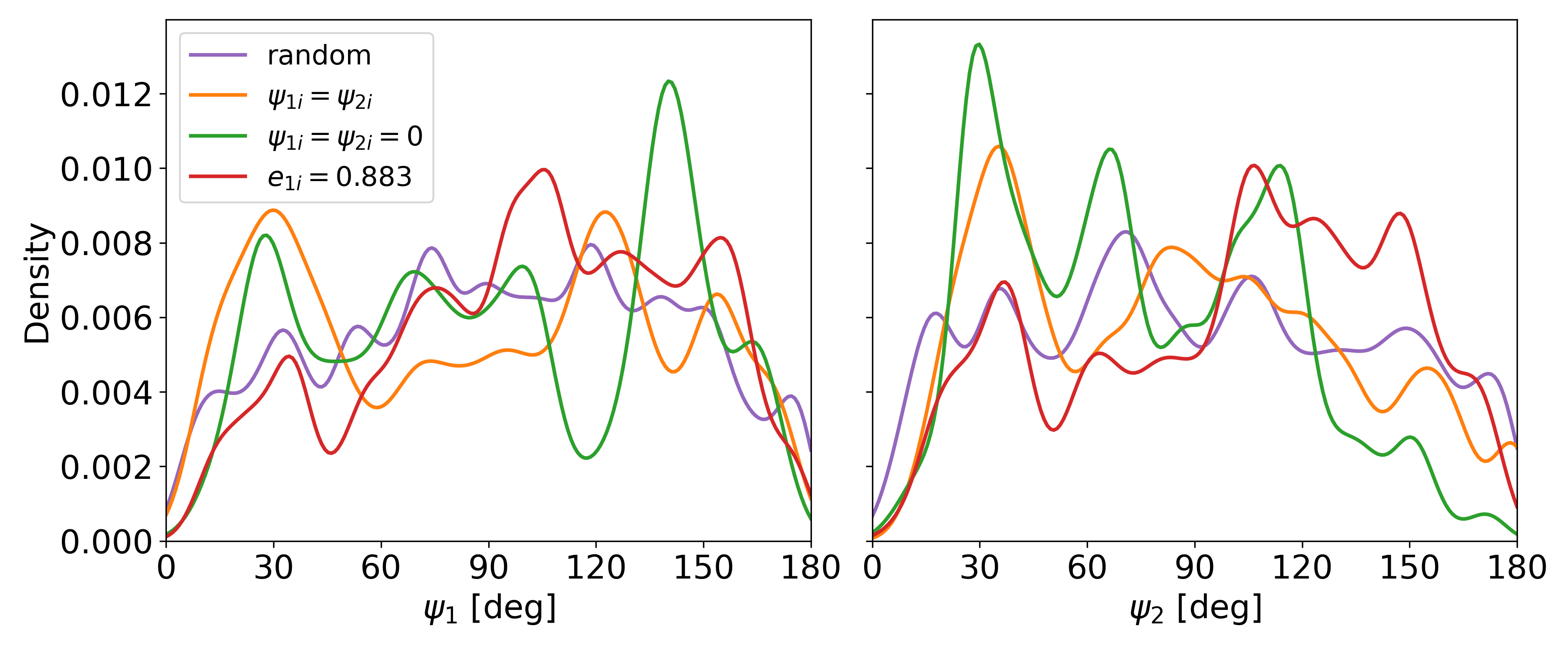}
 \caption{Behavior of $\psi_{1}$ and $\psi_{2}$, the obliquities of VHS 1256 A and VHS 1256 B, respectively, for the systems that evolved to match the observed configuration. The \textbf{left} panel depicts the obliquity distribution for one member of the inner binary, while the \textbf{right} panel depicts the obliquity distribution for the other. In both panels, the purple line corresponds to systems for which $\psi_{1i}$ and $\psi_{2i}$ were independently chosen from uniform distributions between $0$ and $\pi$ and $e_{1i}$ was chosen from a uniform distribution between $0$ and $1$ (i.e., the systems plotted in Figure \ref{fig:timestamps16}.) The orange line corresponds to systems for which $\psi_{1i}$ was chosen from a uniform distribution between $0$ and $\pi$, $\psi_{2i}$ was chosen to be the same as $\psi_{1i}$, and $e_{1i}$ was chosen between $0$ and $1$. The green line corresponds to systems for which $\psi_{1i}$ and $\psi_{2i}$ were both initialized at $0$, and $e_{1i}$ was again chosen between $0$ and $1$. The red line depicts systems where $\psi_{1i}$ and $\psi_{2i}$ were chosen independently between $0$ and $\pi$, and $e_1$ was initialized at $e_{1i}=0.883$. In both panels, all $400$ time points between $120-160$~Myr are considered and accumulated into the pictured density contours. There is no correlation between $\psi_1$ and $\psi_2$.}
 \label{fig:psi1psi2}
 \end{center}
\end{figure*}

%The triple dynamical evolution has a distinct prediction... {\sn{NEED TO ADD ONCE WE HAVE THE INITIAL ZERO BOliquity results - what about the initial eccentric systems, what obliquity distribution do they predict?}}

Figure \ref{fig:psi1psi2} shows the distributions of the stellar obliquities $\psi_{1}$ and $\psi_{2}$ of the inner binary members (VHS 1256 A and VHS 1256 B) for systems that evolve into the observational window at one or more of the $400$ time points between $120-160$~Myr (the estimated age of the system). Each distribution is time-stacked across all snapshots, so the contours reflect both the diversity of configurations reached and the relative amount of time spent near different obliquities. The systems that dynamically evolved into the observational window have a wide variety of spin-orbit angles for the inner binary members. 

Overall, there is not an overwhelmingly preferred $(\psi_{1}, \psi_{2})$ configuration, and there is no correlation between these two angles. The left panel depicts the distribution for $\psi_1$ while the right panel depicts the distribution for $\psi_2$. On both panels, the purple distribution depicts the systems where $\psi_{1i}$ and $\psi_{2i}$ were chosen independently from uniform distributions between $0$ and $\pi$ (ie, the systems plotted in Figure \ref{fig:timestamps16}). The orange distribution represents systems that started with $\psi_{1i}=\psi_{2i}$ with $\psi_{1i}$ again chosen from a uniform distribution between $0$ and $\pi$, and the green distribution depicts systems that started with $\psi_{1i}=\psi_{2i}=0$. These three distributions (ie, purple, orange, and green) all had $e_{1i}$ chosen from a uniform distribution between $0$ and $1$. The red distribution depicts systems that were initialized with $e_{1i}=0.883$, and with $\psi_{1i}$ and $\psi_{2i}$  chosen independently from uniform distributions between $0$ and $\pi$. 

The systems with $\psi_{1i}$ and $\psi_{2i}$ chosen independently and $e_{1i}$ chosen between $0$ and $1$ (ie, the purple systems) have a relatively uniform distribution in the final configuration of $\psi_{1}$ and $\psi_{2}$. Likewise, the systems with $\psi_{1i}=\psi_{2i}$ (ie, the orange systems) show similar behavior with a mostly uniform distribution with small peaks near $\psi_1\sim30^{\circ}$, $\psi_1\sim120^{\circ}$, and $\psi_2\sim30^{\circ}$. The systems with $\psi_{1i}=\psi_{2i}=0$ (ie, the green systems) have a small peak in the distribution near $\psi_1\sim30^{\circ}$, and larger peaks near $\psi_1\sim145^{\circ}$ and $\psi_2\sim30^{\circ}$. The systems initialized with $e_{1i}=0.883$ show a moderate preference for a configuration with $\psi_1\sim\psi_2\sim115^{\circ}$. The behavior of these two angles can be understood as a consequence of the precession of the spin vectors of VHS 1256 A and VHS 1256 B. Under the presence of external torques, an object’s spin axis will naturally precess; our simulations incorporate the spin precession of all three members of this system.

%\begin{figure*}
%  \begin{center} %\vspace{-1.2cm} %\hspace{0.2cm} 
%    \includegraphics[width=0.7\textwidth]{Example2_orbit.png}
%    \par\vspace{1ex}
%    \parbox{0.7\textwidth}{ % Match figure width
%  \caption{  \upshape   An example time evolution for one of the samples with $m_{3}\approx 16M_{J}$ that landed in the target region. We note that the eccentricity was excited from an initial value of 0.195. The inclination was stable except for the four peaks corresponding to the eccentricity excursions.   } \label{fig:Ex2_orbit} 
 % }
 %\end{center}
%\end{figure*}

\section{Possible origins }\label{sec:origins}

\subsection{A hidden fourth companion}\label{ssec:4body}

\begin{figure*}[t]
 \centering
 \includegraphics[width=\textwidth]{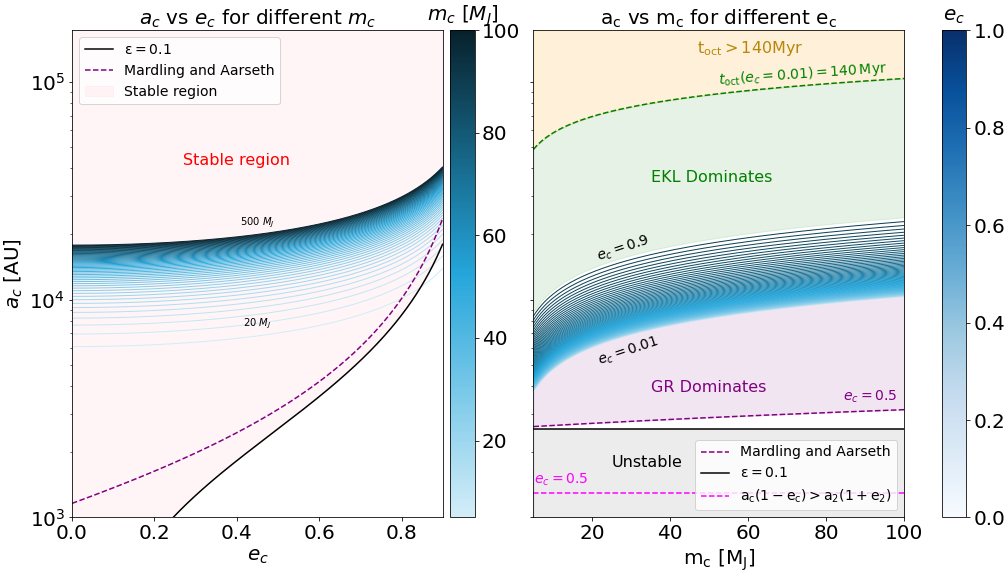}
 \caption{ \textbf{Left:} Stability of a potential fourth companion in the $a_c$–$e_c$ plane. The purple dashed line shows the \citet{Mardling+01} criterion for $m_c=35$~M$_{\rm J}$ and $i_{2,c}=65^{\circ}$. The black solid line shows the $\epsilon$ stability criterion from \citet{Lithwick+11}. If the 1pN precession timescale $t_{\rm 1256\ b}^{\rm 1pN}$ is shorter than the quadrupole timescale $t_{\rm quad}$, eccentricity excitation is suppressed. The blue lines correspond to companion masses from $0$–$100$~M$_{\rm J}$, computed by equating $t_{\rm quad}$ with $t_{\rm 1256\ b}^{\rm 1pN}$. Below each line, GR precession suppresses inclination/eccentricity excitation; above each line, EKL can flip the orbit.
\textbf{Right:} Stability of a potential fourth companion in the $a_c$–$m_c$ plane. The dashed green line is where the octupole timescale equals the system age ($140$~Myr) for $e_c=0.01$. %\sout{; since flipping requires $t_{\rm oct}$ shorter than the system age, and this is a limiting nearly circular case}.
Above this line, $t_{\rm oct}>140$~Myr and flips are not expected. Blue lines show eccentricities $0.01$–$0.9$, set by $t_{\rm quad}=t^{\rm 1pN}_{\rm 1256\ b}$; above these lines, EKL can excite VHS 1256 b's eccentricity and produce flips, while below GR precession suppresses them. The purple dashed line shows the \citet{Mardling+01} criterion for $e_c=0.5$. The black line shows the \citet{Lithwick+11} $\epsilon$ criterion. Stable configurations require the companion's minimum pericenter to exceed the maximum apocenter of VHS 1256 b. %; the pink dashed line marks equality for $e_c=0.5$.
The pink dashed line marks this equality for $e_c=0.5$; below the black line, the system is unstable. We note these parameters fall below the {\it Gaia} exclusion region.
 }
 \label{fig:companion}
\end{figure*}
If the interaction between VHS 1256 AB and VHS 1256 b is insufficient to explain the planet’s current nearly polar retrograde orbit, an alternative possibility is the presence of an additional, more distant companion. To explore this hypothesis, we consider the torque by the fourth, hypothetical companion, with semimajor axis $a_c$ and eccentricity $e_c$ on the orbit of the tertiary, VHS 1256 b. In this scenario, VHS 1256 AB (treated here as a single body) and VHS 1256 b would form the ‘inner binary’ of a hierarchical triple system, with the putative companion acting as the outer body. Here, the assumption is that VHS 1256 b formed in a protoplanetary disk around the AB binary. Thus, we assume the planet initially was in a prograde configuration and was later flipped to a retrograde configuration by a far-away companion. In the following equations, $m_{\rm A}$ and $m_{\rm B}$ correspond to the mass of the inner binary members VHS 1256 A and VHS 1256 B. $m_{b}$ denotes the mass of VHS 1256 b, while $m_{c}$ denotes the mass of the proposed companion. $a_{2}$ and $e_{2}$ are the semimajor axis and eccentricity of VHS 1256 b's orbit.

The dynamical effects of such a configuration, to the lowest order of approximation, can be evaluated using the quadrupole timescale \citep[e.g.,][]{Pejcha+13,Hamers+15,Grishin+18,Klein+25}.   
\begin{equation}
    t_{\rm quad} \sim \frac{2\pi a_{c}^{3}(1-e_{c}^{2})^{3/2}\sqrt{m_{A}+m_{B}+m_{b}}}{a_{2}^{3/2}m_{c}k} \ ,
\end{equation}
where $k^2$ is the gravitational constant.

The inner orbit of VHS 1256 b also experiences GR precession of its pericenter, arising from the first post-Newtonian (1PN) correction to the gravitational potential. The corresponding 1PN precession timescale depends on the speed of light, $c$, because the correction scales as $v^{2}/c^{2}$  in the post-Newtonian expansion \citep[e.g.,][]{Naoz+13b}:
\begin{equation}
    t_{\rm 1256\ b}^{\rm 1pN} \sim 2\pi\frac{a_{2}^{5/2}c^{2}(1-e_{2}^{2})}{3k^{3}(m_{A}+m_{B}+m_{b})^{3/2}} \ . 
\end{equation}
When $t_{\rm 1pN} < t_{\rm quad}$, relativistic precession dominates and suppresses the secular excitation of eccentricity and inclination \citep[e.g.,][]{Naoz+13b, Naoz+20}. Figure \ref{fig:companion} shows the loci where $t_{\rm 1256\,b}^{\rm1pN}=t_{\rm quad}$ in the $a_c-e_c$ (left) and $a_c-m_c$ (right) planes. Above each line, inclination flips are possible, whereas below the line, GR precession quenches the dynamical excitation. 

%to constrain the parameter space of such a companion. If the 1PN timescale is shorter than the quadrupole timescale, eccentricity excitations are suppressed \citep{Naoz+13b, Naoz+20}. Thus, we equate the two to determine the SMA of a companion capable of significantly exciting the eccentricity and flipping the orbit of VHS 1256 b. These results are shown in Figure \ref{fig:companion}. 

To achieve flips, the octupole timescale, $t_{\rm oct}$, should be shorter than the age of the system. The 
%These results highlight the relationship between SMA and mass for a tertiary companion with a range of eccentricities. The green dashed line shows the SMA of a potential companion when the 
octupole timescale \citep[e.g.,][]{Weldon+24} is
\begin{equation}
    t_{\rm oct} \sim\frac{64}{15}\frac{a_{c}^{4}(1-e_{c}^{2})^{5/2}(m_{A}+m_{B}+m_{b})^{3/2}}{a_{2}^{5/2}e_{c}k|m_{A}+m_{B}-m_{b}|m_{c}} \ .
\end{equation}
The result of equating this timescale to the system's age ($140$~Myr) is depicted as the green dashed line in Figure \ref{fig:companion}. 
%is equated with the age of the system. 
The timescale increases with the companion's eccentricity, so to maximally constrain the parameter space for all values of $e_{c}$, we consider here the limiting case of a nearly circular orbit with $e_{c}=0.01$. 

Additionally, the quadruple body system needs to obey long-term stability, at least to preserve the inner triple. We adopt two conservative stability criteria, the first of which is following \citet{Mardling+01},
\begin{equation}
    \frac{a_{c}}{a_{2}}>2.8\bigg(1+\frac{m_{c}}{m_{A}+m_{B}+m_{b}}\bigg)^{2/5}\frac{(1+e_{c})^{2/5}}{(1-e_{c})^{6/5}}\bigg(1-\frac{0.3i_{2,c}}{180}\bigg) \ ,
\end{equation} 
where $i_{2,c}$ denotes the mutual inclination between the orbit of VHS 1256 b and the proposed companion. We show this line as the dashed purple lines, corresponding to the minimum value of $a_{c}$ needed to ensure the stability. The second is using 
\begin{equation}
    \epsilon = \frac{a_2}{a_c}\frac{e_c}{1-e_c^2} \ ,
\end{equation}
\citep[e.g.,][]{Lithwick+11}. $\epsilon$ is the dimensionless prefactor that modulates the strength of the octupole-order term in the hierarchical three body Hamiltonian \citep{Lithwick+11, Naoz+13}. 

%The purple line displays the minimum value of $a_{c}$ needed to ensure the stability of the system using the stability criterion from 

This Figure demonstrates that 
%This curve corresponds to a companion with an eccentricity of $0.5$ and an inclination of $65^{\circ}$, demonstrating that
a companion capable of flipping VHS 1256 b’s orbit could exist in a stable configuration orbiting the triple system. In other words, a companion with a mass of $\geq 10$~M$_{\rm J}$ and a semi-major axis of a few to tens of thousands of au for a range of eccentricities can precipitate a flip to the orbit of the planet. 

Significantly, the parameter space considered in Figure \ref{fig:companion} is limited to the parameter space inaccessible by {\it Gaia}. Specifically, {\it Gaia} would have detected a companion with mass $\geq 100$~M$_{\rm J}$ at separations $\geq 10$~au \citep{GaiaDR3, Perryman+14}. 
Thus, we predict that a companion, in the regimes presented in Figure \ref{fig:companion} can naturally allow for such a flip. 

%We note that this is unlikely given that the parameter space is already highly constrained by GAIA, which rules out a companion with a mass greater than $\sim100M_{J}$.

\begin{figure}[t]
 \centering
 \includegraphics[width=0.5\textwidth]{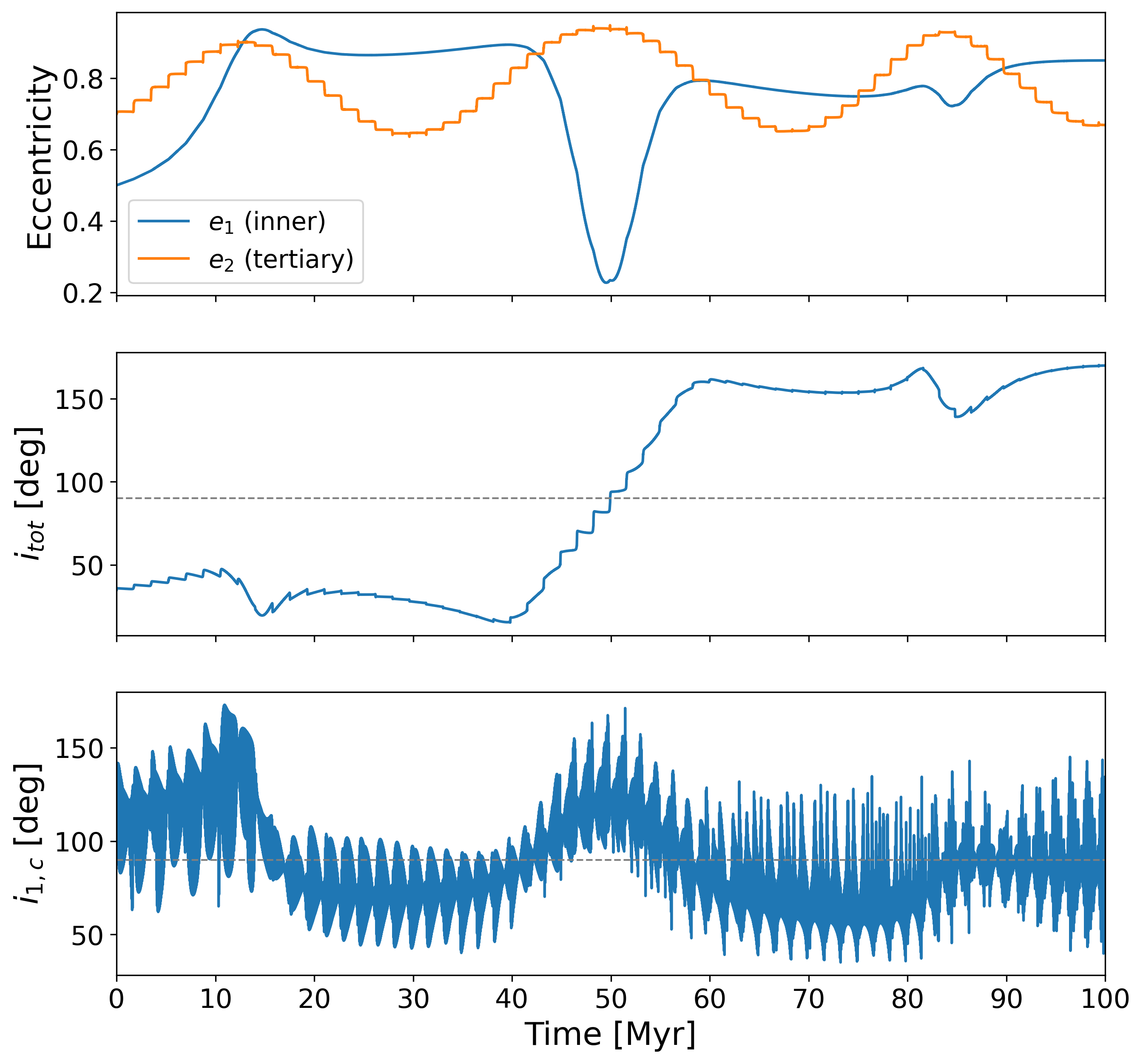}
 \caption{A proof-of-concept 4-body simulation performed with {\tt REBOUND} for a VHS 1256-like system. The proposed companion has a mass $m_{c}=35$~M$_{\rm J}$, semimajor axis $a_{c}=8000$~au, eccentricity $e_{c}=0.84$. The initial inclination between the inner binary and the quaternary companion is $i_{1,c}\sim117^{\circ}$. The semimajor axes of the inner binary and tertiary orbits are $1.96$~au and $350$~au, respectively, and remain stable throughout the duration of the integration and are thus not shown here for clarity. The eccentricities for the inner two orbits are $e_{1}=0.5$ and $e_{2}=0.7$. Such a companion is capable of flipping the initially prograde orbit of VHS 1256 b ($i_{\rm tot}=35^{\circ}$), where $i_{\rm tot}$ is the mutual inclination between the inner binary's orbit and the tertiary's orbit.}
 \label{fig:4body}
\end{figure}

To test this scenario, and to relax the assumption used in this timescale analysis, which  
%This approach 
treats the two inner objects as one, we run 
%and thus does not capture a full dynamical picture of this hypothetical system. In order to better explore this possibility, we examine the 
a 4-body example system using {\tt REBOUND} \citep{rebound, reboundias15}. In this proof-of-concept, we choose a companion with mass $35$~M$_{\rm J}$ with a semimajor axis of $8000$~au and an eccentricity of $e_{c}=0.84$ and evolve the system for $100$~Myr. We initialize VHS 1256 b in a prograde orbit with a mutual inclination between the inner and outer orbits of $i_{\rm tot}=35^{\circ}$. The mutual inclination is calculated by taking the dot product of the two orbital angular momentum vectors. The time evolution of this system is depicted in % results are displayed in 
Figure \ref{fig:4body}. The initial inclination between the VHS 1256 AB binary angular momentum and the quaternary's angular momentum, $i_{1,c}$, is $117^\circ$. Similarly, we compute this mutual inclination by taking the dot product of the relevant orbital angular momentum vectors. Such a large angle may be realized if these systems represent the low-mass tail of wide triple and quadruple systems that are observed to have an isotropic distribution \citep[e.g.,][]{Shariat+25}.  
Note that this example does not include GR precession, which is negligible in this case as highlighted in Figure \ref{fig:4body}. 

For the example presented in Figure \ref{fig:4body}, $i_{\rm tot}$ flipped from prograde to retrograde after about $50$~Myr, and remained this way. Moreover, the eccentricity of the planet and the inner body naturally achieved their observed values. Since the secular approximation remains valid throughout this process\footnote{ In other words, throughout the evolution, the system's semi-major axes of the orbits remain the same}, we expect that the planetary obliquity would reach the observed values as well (as highlighted in Figure \ref{fig:timestamps16}). 

Therefore, we conclude that it is possible that such a companion flipped the orbit of VHS 1256 b, and then EKL, combined with the spin precession, evolved the system into the observed obliquity range. The full numerical analysis of population synthesis of such quadruple systems is beyond the scope of this paper. 
%a combination of EKL, tidal interactions, and general relativistic precession dynamically evolved the system into the observed inclination regime. 
We highlight that this companion falls below the $100$~M$_{\rm J}$ threshold for previous {\it Gaia} detection. {\it Gaia} astrometric measurements can rule out relatively massive companions at small separations. Our proposed companion, with a mass of $35$~M$_{\rm J}$ and a semimajor axis of $8000$~au, falls well below this detection threshold and therefore remains observationally unconstrained.
%For a mass above $100$~M$_J$,  For the minimum mass of $\sim100 M_{J}$, Gaia has excluded a semimajor axis above $\sim10$ AU. 
%Thus, our example companion with a mass of $35$~M$_{J}$ at $8000$~au falls below this threshold. 
Further observations of the VHS 1256 system could better constrain the parameters of such a companion. 

In order to estimate the likelihood that a 4th companion could remain bound, we consider that the inclination for our test case flipped after around $50$~Myr. Thus, we estimate that the system would potentially still be in a dense open cluster birth environment when this flip occurred, given the approximate lifetime of open clusters \citep[e.g.,][]{Kuhn19, Brown+22, Bastian11, Gieles09}. We simulate $200,000$ flyby encounters, selecting our interloper masses from a Kroupa IMF \citep{Kroupa01}. We sample our interloper velocities from a Maxwellian distribution with $\sigma=1$~km~sec$^{-1}$ \citep[e.g.,][]{Kuhn19, Brown+22}, and consider impact parameters up to $3000$~au \citep{Lestrade+11, Pfalzner+21, Proszkow+09}. We find that $\sim40\%$ of the companions remained bound; thus, we conclude that a hidden 4th companion is a plausible origin scenario for the large mutual inclination.

Detecting such a companion would require targeted follow-up observations using complementary techniques. Deep, wide-field direct imaging with instruments such as JWST/NIRCam or Gemini/GPI could potentially reveal a substellar object at thousands of au \citep{Bowler16, Espinoza+23}. Additionally, deep infrared surveys (e.g., VISTA or LSST in the near future) could be used to search for common proper motion companions over multi-year baselines \citep{Cross+12, Ivezi+19}. Because the companion’s gravitational influence is strongest on VHS 1256 b’s wide orbit, long-term astrometric monitoring of the VHS 1256 system may reveal subtle accelerations in the motion of VHS 1256 b relative to the inner binary \citep[e.g.,][]{Lacour+21}. High-precision astrometry from facilities like VLTI/GRAVITY or future {\it Gaia} data releases could place tighter constraints on this effect \citep{Gravity17}.

While our proof-of-concept 4-body simulation reproduced the observed mutual inclination, we note that filament fragmentation is also consistent with the source of the observed misalignment. While all four bodies could have formed via core/filament fragmentation, with the tertiary originally prograde and later flipped to retrograde by the companion, it is also possible for this misalignment to be the result of the randomization of the angular momentum vectors associated with core collapse. Formation of the tertiary in an aligned proto-planetary disk, however, has the limitation that the fourth companion is required in order to explain the observations.

\subsection{A possible flyby}\label{ssec:flyby}
To further explore the dynamical origin of the large mutual inclination, we next consider the possibility that VHS 1256 b started in a prograde configuration with respect to the inner orbit and was flipped by a flyby encounter with a massive body at some point in the system’s history \citep[e.g.,][]{Rodet+22, Brown+22, poon+24, Michaely+16, Michaely+19, Michaely+20, Kaib+14, Hamers+19, shariat+23}. To sample the masses of the potential interlopers, we use a Kroupa IMF \citep{Kroupa01} to reflect the stellar population in an open cluster birth environment. To sample a kicker velocity, we use a Maxwellian distribution with $\sigma=1$~km~sec$^{-1}$ \citep[e.g.,][]{Kuhn19} to reflect the fact that the VHS system is very young and thus was likely still in its birth environment when the proposed kick occurred \citep{Stone+16, poon+24, Dupuy+23}. Therefore, we expect the relative velocity of the interlopers inducing the kicks to be fairly small on the order of a few km~sec$^{-1}$ \citep[e.g.,][]{Kuhn19, Brown+22}. 

Because most stars form in clustered environments that expand and dissolve over tens to hundreds of Myr, with many systems already super-virial by a few Myr, we assume VHS 1256 remained in its birth cluster/association for $\sim10-100$~Myr \citep[e.g.,][]{Kuhn19, Brown+22, Bastian11, Gieles09}. We sample impact parameters in the range $b=800-3000$~au. The lower limit of $800$~au corresponds to close stellar passages that are strong enough to impart a significant velocity kick to a wide companion like VHS 1256 b, potentially exciting its mutual inclination or even unbinding it. The upper limit of $3000$~au reflects the scale where the interaction cross section remains appreciable in typical open cluster environments, but encounters beyond this distance produce largely negligible perturbations \citep[e.g.,][]{Lestrade+11}. This range therefore brackets the regime where most dynamically relevant encounters are expected to occur over $10-100$~Myr in open-cluster or association environments \citep[e.g.,][]{Lestrade+11, Pfalzner+21, Proszkow+09}. We choose $i_{\rm tot,i}$ uniformly between $0^{\circ}-89^{\circ}$ to ensure that the system is in a prograde configuration prior to the kick. The rest of the orbital parameters are chosen to be consistent with our overall population, as described in Section \ref{sec:triple}. $e_{1}$ and $e_{2}$ are chosen from uniform distributions between $0$ and $1$ and $0.6$ and $0.8$, respectively. Lastly, $g_{1}$ and $g_{2}$ are drawn from uniform distributions from $0$ to $2\pi$. We set $a_{1}=1.96$~au and sample $a_{2}$ from a uniform distribution between $200$ and $460$~au. We set $m_{1}=0.072$~M$_{\odot}$, $m_{2}=0.069$~M$_{\odot}$, and $m_{3}=0.015$~M$_{\odot}$. %The physical motivations for these choices mirror those described in Section \ref{sec:triple}.

%We simulate the kick of $50,000$ of triple systems and then examine the final configuration, assessing if the orbits remained bound and if the system flipped or not.
We simulate the kick of 50,000 triple systems and examine the final configuration, assessing whether the orbits remained bound and if the system flipped.
The post kick orbital configuration is calculated following \citet{Lu+19}. % We follow the prescription presented in \citet{Lu+19} to calculate the orbital elements post-kick and present these results in Figure \ref{fig:flyby}. 
In particular, we are interested in systems that post-kick flipped, because this will yield the initial conditions needed to form a planet with the observed inclination configuration (see Figure \ref{fig:timestamps16}). 
%The blue line in Figure \ref{fig:flyby} represents the systems that flipped. %We highlight that from the three body integration, presented in Section \ref{sec:triple}, (see Figure \ref{fig:timestamps16}), the aim here is to achieve a retrograde configuration, which then naturally can lead to the observed architecture of the system.

\begin{figure}
  \begin{center} %\vspace{-1.2cm} %\hspace{0.2cm} 
    \includegraphics[width=\linewidth]{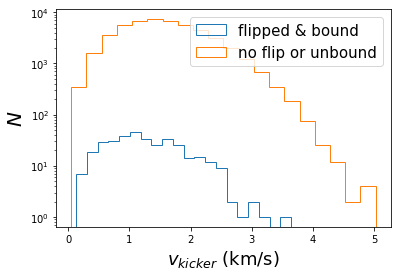}
  \end{center} %\vspace{-0.4cm} %\hspace{0.2cm}
  \caption{  \upshape The blue line depicts the $343$ systems out of the $50,000$ sample systems that stayed bound and flipped after an impulsive kick to the system. The orange line depicts the systems that either became unbound, did not flip their inclination into the observed regime, or both. We highlight that most of the realizations (96\%) remain bound. The x-axis represents the velocity of the interloper responsible for imparting the kick.   } \label{fig:flyby} %\vspace{-0.4cm}
\end{figure}

We find that this is extremely rare, as only $\sim0.7\%$ of the systems survived and flipped into the observed inclination regime. These results assume the kick was imparted on the tertiary; however, applying the kick to the inner binary produces the same success rate of $\sim0.7\%$. These systems had considerably slower kicker velocities between $\sim0.1$ and $\sim 3$~km~sec$^{-1}$ and are depicted in the blue line in Figure \ref{fig:flyby}. The rest of the systems, i.e., those that 
%The orange line on this histogram shows the number of systems that 
either did not flip, had one or both of the orbits unbind, or both, are shown in the orange line. 

%. The blue line represents the systems that flipped into the observed inclination regime. For kicker velocities between $\sim0.1$ and $\sim 7.0 \text{ km/s}$, resulting in kicks approximately $0.1-2.0$ km/s,  $102$ of the original $50,000$ triple systems ($\sim0.2\%$) survived and flipped. We thus conclude that slower interlopers have a greater likelihood of flipping VHS 1256 b's orbit, consistent with interactions within the birth cluster of the system. 

In order to gauge the probability that a flyby induced a successful kick, we first assume a density of $n=100~\text{pc}^{-3}$ \citep{Brown+22}. We again sample the relative velocity from a Maxwellian with $\sigma=1$~km~sec$^{-1}$ \citep[e.g.,][]{Brown+22, Kuhn19} and set the impact parameter to be $b=800-3000$~au and use this to compute the cross section, accounting for gravitational focusing. For the mass of the flyby perturber, we again use a Kroupa IMF \citep{Kroupa01}. We use the impulse approximation \citep[e.g.,][]{Kaib+14} to compute $\Delta v$ via
\begin{equation}
    \Delta v\approx\frac{2GM}{bv_{rel}} \ ,
\end{equation}
where $M$ is the mass of the perturber, $b$ is the impact parameter, and $v_{rel}$ is the relative velocity sampled from the Maxwellian. We estimate the average time between successful flybys with
\begin{equation}
    t_{\rm flyby} = \frac{1}{n\sigma_{\rm cs} v_{\rm rel}} \ ,
\end{equation}
\citep[e.g.,][]{Binney+08} where $\sigma_{\rm cs}$ is the encounter cross section and estimate that a perturber capable of sufficiently exciting the inclination interacts with the VHS 1256 system every $\sim2.9\times10^{9}$~yr.

The velocity of the perturber here was taken to reflect a dense environment such as a star association or a star cluster. We note that clusters often disperse within a few tens of Myrs \citep[e.g.,][]{Bastian11, Gieles09}, so the density and velocity estimates are likely only valid for the beginning of the system's lifetime. Thus, consistently with \citet{poon+24}, we find that the observed inclination cannot be well explained by a flyby encounter. Given the system's age and likely dispersal of its birth cluster, the probability of such a strong encounter occurring within its lifetime is extremely small. A full analysis of potential flyby interactions the VHS 1256 system could have experienced after leaving the birth cluster environment is beyond the scope of this paper.

\subsection{Filament Fragmentation}\label{ssec:filament}
It suggested that fragmentation can readily produce wide, roughly coeval substellar multiples with separations of hundreds to thousands of au and a broad range of mutual inclinations \citep[e.g.,][]{Offner+16,Kratter10,Lee+19}.  In this context, VHS 1256 b’s wide separation ($\approx400$~au) and near-polar orientation fall within the diversity of outcomes expected from turbulent core fragmentation. However, testing this pathway as a formation scenario requires population-level statistics (e.g., determining whether VHS 1256-like systems follow a broader trend of polar or misaligned triples consistent with turbulent fragmentation).

\section{discussion}\label{sec:discussion} 
Circumbinary planets (CBPs) occupy a distinctive class in planetary science because they form and evolve in a gravitational landscape far more complex than that of planets around single stars \citep[e.g.,][]{Welsh+12, Orosz+12N, Doyle+11}. Their orbital properties record a dynamical history shaped by the combined influence of two stellar hosts \citep[e.g.,][]{Artymowicz+94, Paardekooper+08}, the natal disk, and, in some cases, additional companions or external perturbations \citep[e.g.][]{Schwarz16, Martin+15, Martin17}.
While most CBPs, detected in transit surveys, are coplanar, this is clearly a selection eﬀect \citep[e.g.,][]{Li+16, Martin+15, Pierens18, Goldberg23}. 

A few misaligned circumbinary planets (such as Kepler-413b \citep{Kostov+14}, Kepler-453b \citep{Welsh+12, Welsh+15}, 2M1510b \citep{Baycroft+25}, and VHS 1256 b \citep{poon+24, Stone+16, Dupuy+23}) have already been detected. These misaligned CBPs challenge standard formation scenarios and provide valuable insight into processes such as multi-body dynamics, secular perturbations, and dynamical interactions with additional bodies or passing stars \citep{Michaely+14, Kostov+14, Welsh+15, Kaib+13, Malmberg+11}. The VHS 1256-1257 system explored here, with its retrograde, nearly polar tertiary orbiting an extremely low-mass binary, is a striking example of such complexity and offers an opportunity to explore the interplay of multiple dynamical mechanisms in sculpting extreme planetary architectures.

To investigate whether the observed orbital properties of VHS 1256 could arise from hierarchical triple dynamics, we modeled the system using secular three-body dynamics expanded to hexadecapole order \citep[e.g.,][]{Will+17}. This high-order treatment captures the subtle dynamical effects that become important in systems with comparable masses for the inner orbit, and it allows for accurate tracking of the coupled evolution of the inner binary and the tertiary’s orbit. Our simulations indicate that triple-induced perturbations can naturally excite the large eccentricity of the inner binary and drive the planet’s obliquity toward a near-polar configuration relative to the outer binary’s angular momentum (as highlighted Figures \ref{fig:Ex1_orbit}, and \ref{fig:timestamps16}). Particularly, a wide range of initial conditions of the inner binary eccentricity and the planet's obliquity can readily achieve the observed architecture. 

However, triple dynamics alone cannot generate the extreme mutual inclination, both near-polar and retrograde, observed between the inner binary and the planet’s orbit. While a small fraction of systems do undergo orbital flips (see the bottom-left panel in Figure \ref{fig:timestamps16} and Figure \ref{fig:timestamps12} in Appendix \ref{appendix:m312}), they do not reach the measured mutual inclination of $118^{+12^{\circ}}_{-16^\circ}$. This indicates that an additional perturbative mechanism is required.

%Specifically, while we find that a few systems flip (see the bottom-left panel in Figure \ref{fig:timestamps16} and Figure \ref{fig:timestamps12} in Appendix \ref{appendix:m312}), it does not reach the observed mutual inclination of $118^{+12^{\circ}}_{-16^\circ}$. Therefore, our simulations indicate that an additional perturbative mechanism must be at work. 

We therefore explored two external sources of inclination excitation: a hidden fourth companion and stellar flybys. A proof-of-concept four-body integration using {\tt REBOUND} demonstrates that an undetected companion could indeed raise the mutual inclination to the observed values (see Figure \ref{fig:4body}). A plausible configuration is a $\sim35$~M$_{\rm J}$ object at a semimajor axis of $\sim8000$~au, well below {\it Gaia}’s current $\sim100$~M$_{\rm J}$ detection threshold for such wide orbits, and therefore observationally unconstrained (see Figure \ref{fig:companion}). Our flyby simulations show that $\sim40\%$ of companions in this regime would remain bound after cluster dispersal, making this a dynamically viable origin scenario. Detecting such a companion would require targeted follow-up, such as deep wide-field imaging (e.g., JWST/NIRCam, Gemini/GPI), multi-epoch infrared surveys (e.g., VISTA, LSST), or long-term astrometric monitoring (e.g., VLTI/GRAVITY, future {\it Gaia} releases), which could reveal either the companion directly or its subtle gravitational influence on VHS 1256 b’s wide orbit. By contrast, we find that the probability of a sufficiently strong stellar flyby producing the observed inclination is exceedingly small (Figure \ref{fig:flyby}), consistent with the results of \citet{poon+24}. This rules out flybys as a likely origin pathway for VHS 1256.

Both the fly-by and fourth-companion scenarios rely on a planet-formation pathway in which VHS 1256b formed within a circumbinary disk around the inner binary, analogous to planet formation around single stars. Several directly imaged systems, such as HR 8799 and AB Aurigae, host giant planets ($\sim7$~M${\rm J}$, $9-20$~M$_{\rm J}$) at large separations of $\sim$70–100~au from their host stars \citep{Marois+08, Shibaike+25}, and these systems have been suggested to possess extended protoplanetary disks reaching several hundred au \citep{Booth+16, Boccaletti+20}.  Furthermore, the YSES 1 system hosts two giant exoplanets ($6 \pm 1$~M$_J$, $14 \pm 3$~M$_J$) on very wide orbits ($160, 320$~au) \citep{Bohn+20}. Thus, while a planet at $\sim400$~au from its host binary may at first glance appear extreme, VHS 1256 b lies within the regime of other wide-orbit, directly imaged planets. Yet, forming such an extended and massive protoplanetary disk around such a low-mass binary ($\sim0.14$~M${\odot}$) poses a significant challenge. Observed disks around single late-M or substellar hosts are typically compact, with radii of only $\sim100–200$~au \citep[e.g.,][]{Zhang+25, Long+18} and forming a $\gtrsim10–15$~M$_{\rm J}$ companion in situ at $\sim400$~au would require an unusually large and massive disk. Further, current formation models as well as observational ALMA surveys indicate that binaries show systematically depleted dust reservoirs compared to single stars, consistent with tidal truncation and dynamical stirring by the binary \citep{Andrews+18, Mayer+05, Flaherty+25}.  Nonetheless, if such a disk did exist, previous studies \citep[e.g.,][]{Martin17,Chen_Martin+20,Martin18,Chen_Martin+21,Martin+23,Martin+25,Martin+19} suggest that a polar configuration could naturally emerge, potentially explaining the current orientation of VHS 1256b’s orbit.  Furthermore, another possibility is that the system had a closer-in planet that was ejected early in the system’s lifetime. In this scenario, VHS 1256 b could have formed with a more modest separation from the inner binary and was pulled onto a very wide, eccentric orbit following the ejection \citep[e.g.,][]{Chen+22}. Another potential origin for the wide tertiary is dynamical capture. \citet{Li+16} explored the capture of Planet Nine-like bodies during the Sun’s residence in its birth cluster and found that successful capture requires unusually low relative velocities ($\geq1$~km~s$^-1$) and close encounters within a few hundred au. Their estimated capture probability is $\geq1$\% for Solar-type stars, even in dense clusters. Given the lower total mass of the VHS 1256–1257 binary ($\sim0.14$~M$_{\odot}$) and its modest binding energy ($\sim1.8\times10^5$~J kg$^-1$ at $350$~au), the likelihood of retaining a captured $\sim15$~M$_J$ object at $\geq300$~au is even smaller (\cite{Li+16}).

Alternatively, formation through core or filament fragmentation \citep[e.g.,][]{Offner+16,Offner+22,Lee+19,Bate+02,Bate09,Chabrier+14,Tokovinin17} can naturally produce a randomized set of angular momentum vectors, offering a straightforward explanation for the extreme mutual inclination observed in VHS 1256 b \citep{poon+24}. Confirming this scenario will require population-level statistics to assess whether VHS 1256–like systems follow a broader trend of polar or misaligned triples consistent with turbulent fragmentation. In contrast, the late dynamical excitation pathway we explore here, via a fourth companion or stellar flyby, is more readily testable in individual systems. The existence of such a companion could, in principle, be verified through future high-precision astrometric measurements (e.g., Gaia DR4). Motivated by this, we investigate whether such perturbations can reproduce VHS 1256 b’s present-day obliquity and mutual inclination.

\section{acknowledgement}
 We would also like to thank the anonymous referee, Gongjie Li, Rebecca Martin, Michael Poon, and Hannao Rein for the fruitful discussions and comments.
LH and SN thank Howard and Astrid Preston for their generous support. CS acknowledges support from the Department of Energy Computational Science Graduate Fellowship supported by the U.S. Department of Energy, Office of Science, Office of Advanced Scientific Computing Research, under Award Number DE-SC0026073.
We would like to thank UCLA Office of Advanced Research Computing's Research Technology group for the computational and storage services provided by the Hoffman2 Shared Cluster, and the UCLA department of physics and astronomy for their continued support. 

%appendix 
\clearpage
\FloatBarrier
\appendix
\section{Quadrupole Equations of Motion}\label{appendix:quad}

\renewcommand{\theequation}{A\arabic{equation}}
\setcounter{equation}{0}
\noindent
To investigate the dynamical behavior of the system, we perturbatively expand the Hamiltonian,
\begin{equation}
    H=\frac{k^{2}m_{1}m_{2}}{2a_{1}}+\frac{k^{2}m_{3}(m_{1}+m_{2})}{2a_{2}}+\frac{k^{2}}{a_{2}}\sum_{n=2}^{\infty}\Big(\frac{a_{1}}{a_{2}}\Big)^{n}M_{n}\Big(\frac{r_{1}}{a_{1}}\Big)^{n}\Big(\frac{a_{2}}{r_{2}}\Big)^{n+1}P_{n}(\cos\Phi)
\end{equation}
where
\begin{equation}
    M_{n}=m_{1}m_{2}m_{3}\frac{m_{1}^{n-1}-(-m_{2})^{n-1}}{(m_{1}+m_{2})^{n}}
\end{equation}
to hexadecapole order and then average the Hamiltonian via the execution of two Von-Zeipel transformations (one for each orbit). We then integrate the equations of motion derived from our double-averaged Hamiltonian up to the hexadecapole level of approximation; see Appendix B for the octupole order term in the equation sof motion and Appendix C for the hexadecapole order term in the equations of motion.
\begin{equation}
    L_{1} = \frac{m_{1} m_{2}}{m_{1} + m_{2}} \sqrt{k^{2} (m_{1} + m_{2}) a_{1}}
\end{equation}

\begin{equation}
    L_{2} = \frac{m_{3} (m_{1} + m_{2})}{m_{1} + m_{2} + m_{3}} \sqrt{k^{2} (m_{1} + m_{2} + m_{3}) a_{2}}
\end{equation}

\begin{equation}
    G_{1} = L_{1} \sqrt{1 - e_{1}^{2}}, \qquad
    G_{2} = L_{2} \sqrt{1 - e_{2}^{2}}
\end{equation}

\begin{equation}
    C_{2} = \frac{k^{4}}{16} \frac{(m_{1} + m_{2})^{6}}{(m_{1} + m_{2} + m_{3})^{3}} 
    \frac{m_{3}^{7}}{(m_{1} m_{2})^{3}} \frac{L_{1}^{4}}{L_{2}^{3} G_{2}^{3}}
\end{equation}

\vspace{1em}
\noindent \textbf{Eccentricity evolution:}

\begin{equation}
    \left. \frac{de_{1}}{dt} \right|_{\text{Quad}} = 
    C_{2} \frac{1 - e_{1}^{2}}{G_{1}} 
    \cdot 30 e_{1} \sin^{2}(i_{\text{tot}}) \sin(2g_{1})
\end{equation}

\begin{equation}
    \left. \frac{de_{2}}{dt} \right|_{\text{Quad}} = 0
\end{equation}

\vspace{1em}
\noindent \textbf{Argument of periapse evolution:}

\begin{align}
    \left. \frac{dg_{1}}{dt} \right|_{\text{Quad}} = 6 C_{2} \bigg\{ 
    & \frac{1}{G_{2}} \Big[ 4 \cos^{2}(i_{\text{tot}}) 
    + \big(5 \cos(2g_{1}) - 1\big) \big(1 - e_{1}^{2} - \cos^{2}(i_{\text{tot}}) \big) \Big] \nonumber \\
    & + \frac{\cos(i_{\text{tot}})}{G_{2}} \big[2 + e_{1}^{2} (3 - 5 \cos(2g_{1})) \big]
    \bigg\}
\end{align}

\begin{align}
    \left. \frac{dg_{2}}{dt} \right|_{\text{Quad}} = 3 C_{2} \bigg\{
    & \frac{2 \cos(i_{\text{tot}})}{G_{1}} \Big[ 2 + e_{1}^{2} (3 - 5 \cos(2g_{1})) \Big] \nonumber \\
    & + \frac{1}{G_{2}} \Big[ 4 + 6 e_{1}^{2} 
    + \big(5 \cos^{2}(i_{\text{tot}}) - 3 \big) 
      \big(2 + e_{1}^{2} (3 - 5 \cos(2g_{1})) \big) \Big] 
    \bigg\}
\end{align}

\vspace{1em}
\noindent \textbf{Angular momentum evolution:}

\begin{equation}
    \left. \frac{dG_{1}}{dt} \right|_{\text{Quad}} = 
    -30 C_{2} e_{1}^{2} \sin(i_{2}) \sin(i_{\text{tot}}) \sin(2g_{1})
\end{equation}

\begin{equation}
    \left. \frac{dG_{2}}{dt} \right|_{\text{Quad}} = 0
\end{equation}
\\
\vspace{1em}
\noindent \textbf{Longitude of ascending node:}

\begin{equation}
    \left. \frac{dh_{1}}{dt} \right|_{\text{Quad}} = 
    -\frac{3 C_{2}}{G_{1} \sin(i_{1})}
    \big( 2 + 3 e_{1}^{2} - 5 e_{1}^{2} \cos(2g_{1}) \big) \sin(2i_{\text{tot}})
\end{equation}

%\clearpage
\section{Octupole Equations of Motion}\label{appendix:Oct}

\renewcommand{\theequation}{B\arabic{equation}}
\setcounter{equation}{0}

\vspace{1em}
\noindent \textbf{Eccentricity evolution:}

\begin{align}
\left.\frac{de_1}{dt}\right|_{\text{Oct}} = C_{3} e_{2} \frac{1 - e_{1}^{2}}{G_{1}} \bigg[
    & 35 \cos\phi \sin^{2}(i_{\text{tot}}) e_{1}^{2} \sin(2g_{1}) \nonumber \\
    & - 10 \cos(i_{\text{tot}}) \sin^{2}(i_{\text{tot}}) \cos(g_{1}) \sin(g_{2}) (1 - e_{1}^{2}) \nonumber \\
    & - A\big( \sin(g_{1}) \cos(g_{2}) - \cos(i_{\text{tot}}) \cos(g_{1}) \sin(g_{2}) \big)
\bigg]
\end{align}

\begin{align}
\left.\frac{de_{2}}{dt}\right|_{\text{Oct}} = 
- C_{3} e_{1} \frac{1 - e_{2}^{2}}{G_{2}} \bigg[
    & 10 \cos(i_{\text{tot}}) \sin^{2}(i_{\text{tot}}) (1 - e_{1}^{2}) \sin(g_{1}) \cos(g_{2}) \nonumber \\
    & + A\big( \cos(g_{1}) \sin(g_{2}) - \cos(i_{\text{tot}}) \sin(g_{1}) \cos(g_{2}) \big)
\bigg]
\end{align}

\vspace{1em}
\noindent \textbf{Argument of periapse evolution:}

\begin{align}
\left.\frac{dg_{1}}{dt}\right|_{\text{Oct}} = -C_{3} e_{2} \bigg\{ 
    & e_{1} \left( \frac{1}{G_{2}} + \frac{\cos(i_{\text{tot}})}{G_{1}} \right) 
    \bigg[ \sin(g_{1}) \sin(g_{2}) \Big( 10(3\cos^{2}(i_{\text{tot}}) - 1)(1 - e_{1}^{2}) + A \Big) \nonumber \\
    & \hspace{6em} - 5B \cos(i_{\text{tot}}) \cos\phi \bigg] \nonumber \\
    & - \frac{1 - e_{1}^{2}}{e_{1} G_{1}} \bigg[ 
        \sin(g_{1}) \sin(g_{2}) \Big( 10 \cos(i_{\text{tot}}) \sin^{2}(i_{\text{tot}})(1 - 3e_{1}^{2}) \Big) \nonumber \\
    & \hspace{8em} + \cos\phi \big(3A - 10\cos^{2}(i_{\text{tot}}) + 2\big)
    \bigg] 
\bigg\}
\end{align}

\begin{align}
\left.\frac{dg_2}{dt}\right|_{\text{Oct}} = C_{3} e_{1} \bigg\{
    & \sin(g_{1}) \sin(g_{2}) \bigg[
        \left( \frac{4e_{2}^{2} + 1}{e_{2} G_{2}} \right)
        10 \cos(i_{\text{tot}}) \sin^{2}(i_{\text{tot}}) (1 - e_{1}^{2}) \nonumber \\
    & \hspace{5em}
        - e_{2} \left( \frac{1}{G_{1}} + \frac{\cos(i_{\text{tot}})}{G_{2}} \right)
        \bigg( A + 10(3\cos^{2}(i_{\text{tot}}) - 1)(1 - e_{1}^{2}) \bigg)
    \bigg] \nonumber \\
    & + \cos\phi \bigg[
        5B \cos(i_{\text{tot}}) e_{2} \left( \frac{1}{G_{1}} + \frac{\cos(i_{\text{tot}})}{G_{2}} \right)
        + \left( \frac{4e_{2}^{2} + 1}{e_{2} G_{2}} \right) A
    \bigg]
\bigg\}
\end{align}

\vspace{1em}
\noindent \textbf{Angular momentum evolution:}

\begin{align}
\left.\frac{dG_1}{dt}\right|_{\text{Oct}} = C_{3} e_{1} e_{2} \bigg[
    & -35 e_{1}^{2} \sin^{2}(i_{\text{tot}}) \sin(2g_{1}) \cos\phi \nonumber \\
    & + A \big( \sin(g_{1}) \cos(g_{2}) - \cos(i_{\text{tot}}) \cos(g_{1}) \sin(g_{2}) \big) \nonumber \\
    & + 10 \cos(i_{\text{tot}}) \sin^{2}(i_{\text{tot}}) (1 - e_{1}^{2}) \cos(g_{1}) \sin(g_{2})
\bigg]
\end{align}

\begin{align}
\left.\frac{dG_2}{dt}\right|_{\text{Oct}} = C_{3} e_{1} e_{2} \bigg[
    & A \big( \cos(g_{1}) \sin(g_{2}) - \cos(i_{\text{tot}}) \sin(g_{1}) \cos(g_{2}) \big) \nonumber \\
    & + 10 \cos(i_{\text{tot}}) \sin^{2}(i_{\text{tot}}) (1 - e_{1}^{2}) \sin(g_{1}) \cos(g_{2})
\bigg]
\end{align}

\vspace{1em}
\noindent \textbf{Longitude of ascending node:}

\begin{align}
\left.\frac{dh_1}{dt}\right|_{\text{Oct}} = 
- C_{3} e_{1} e_{2} \bigg[
    & 5B \cos(i_{\text{tot}}) \cos\phi 
    - A \sin(g_{1}) \sin(g_{2}) \nonumber \\
    & + 10 \big(1 - 3\cos^{2}(i_{\text{tot}})\big)(1 - e_{1}^{2}) \sin(g_{1}) \sin(g_{2})
\bigg] 
\frac{\sin(i_{\text{tot}})}{G_{1} \sin(i_{1})}
\end{align}
\section{Hexadecapole Equations of Motion}\label{appendix:hex}

\renewcommand{\theequation}{C\arabic{equation}}
\setcounter{equation}{0}
\noindent The following equations are adapted from \cite{Will+17} \\ \\
\noindent \textbf{Eccentricity evolution:}
\begin{tightdisplay}
\begin{align}
\left.\frac{de_1}{dt}\right|_{\text{Hex}} ={}&
\frac{1}{2\pi} \left( \frac{k_2(m_1+m_2)}{a_1^3} \right)^{1/2}
\left( -\frac{315\pi}{1024} \right)
\frac{m_3}{m_1+m_2}
\left( \frac{a_1}{a_2} \right)^5
\left( 1-3\frac{m_1m_2}{(m_1+m_2)^2} \right) \nonumber \\
&\times e_1 \sqrt{1-e_1^2}(1-e_2^2)^{-7/2} \nonumber \\
&\times \Bigg\{
(2+3e_2^2)(1-\cos^2 i_{\mathrm{tot}})
\left[(4+2e_1^2)(1-7\cos^2 i_{\mathrm{tot}})\sin(2g_1)
-21e_1^2(1-\cos^2 i_{\mathrm{tot}})\sin(4g_1)\right] \nonumber \\
& - e_2^2(4+2e_1^2)
\left[(1+\cos i_{\mathrm{tot}})^2(1-7\cos i_{\mathrm{tot}}+7\cos^2 i_{\mathrm{tot}})\sin(2g_1-2g_2)
+ (1-\cos i_{\mathrm{tot}})^2(1+7\cos i_{\mathrm{tot}}+7\cos^2 i_{\mathrm{tot}})\sin(2g_1+2g_2)\right] \nonumber \\
& - 21e_1^2(1-\cos^2 i_{\mathrm{tot}})
\left[(1+\cos i_{\mathrm{tot}})^2\sin(4g_1-2g_2)
+ (1-\cos i_{\mathrm{tot}})^2\sin(4g_1+2g_2)\right]
\Bigg\}
\end{align}
\end{tightdisplay}

\begin{tightdisplay}
\begin{align}
\left.\frac{de_2}{dt}\right|_{\text{Hex}} ={}&
\frac{1}{2\pi} \left( \frac{k_2(m_1+m_2)}{a_1^3} \right)^{1/2}
\left( -\frac{45\pi}{2048} \right)
\frac{m_1m_2}{(m_1+m_2)^2}
\left( 1 - 3\frac{m_1m_2}{(m_1+m_2)^2} \right)
\sqrt{1+\frac{m_3}{m_1+m_2}}
\left( \frac{a_1}{a_2} \right)^{11/2}
\frac{e_2}{(1-e_2^2)^3} \nonumber \\
&\times \Bigg\{
2(8+40e_1^2+15e_1^4)(1-\cos^2 i_{\mathrm{tot}})(1-7\cos^2 i_{\mathrm{tot}})\sin(2g_2) \nonumber \\
& + 28e_1^2(2+e_1^2)
\left[ (1+\cos i_{\mathrm{tot}})^2(1-7\cos i_{\mathrm{tot}}+7\cos^2 i_{\mathrm{tot}})\sin(2g_1-2g_2)
- (1-\cos i_{\mathrm{tot}})^2(1+7\cos i_{\mathrm{tot}}+7\cos^2 i_{\mathrm{tot}})\sin(2g_1+2g_2) \right] \nonumber \\
& + 147e_1^4(1-\cos^2 i_{\mathrm{tot}})
\left[ (1+\cos i_{\mathrm{tot}})^2\sin(4g_1-2g_2)
- (1-\cos i_{\mathrm{tot}})^2\sin(4g_1+2g_2) \right]
\Bigg\}
\end{align}
\end{tightdisplay}

\noindent \textbf{Argument of periapsis evolution:}
\begin{tightdisplay}
\begin{align}
\left.\frac{dg_1}{dt}\right|_{\text{Hex}} ={}&
\frac{1}{2\pi} \left( \frac{k_2(m_1+m_2)}{a_1^3} \right)^{1/2}
\left( \frac{45\pi}{1024} \right)
\frac{m_3}{m_1+m_2}
\left( \frac{a_1}{a_2} \right)^5
\left( 1-3\frac{m_1m_2}{(m_1+m_2)^2} \right)
\frac{\sqrt{1-e_1^2}}{(1-e_2^2)^{7/2}} \times \Bigg\{ \nonumber \\
& (2+3e_2^2) \left[(4+3e_1^2)(3-30\cos^2 i_{\mathrm{tot}}+35\cos^4 i_{\mathrm{tot}})
-28(1+e_1^2)(1-\cos^2 i_{\mathrm{tot}})(1-7\cos^2 i_{\mathrm{tot}})\cos(2g_1) 
+147e_1^2(1-\cos^2 i_{\mathrm{tot}})^2\cos(4g_1)\right] \nonumber \\
& -10e_2^2(4+3e_1^2)(1-\cos^2 i_{\mathrm{tot}})(1-7\cos^2 i_{\mathrm{tot}})\cos(2g_2) \nonumber \\
& +7e_2^2 \Bigg[ 4(1+e_1^2)\left[(1+\cos i_{\mathrm{tot}})^2(1-7\cos i_{\mathrm{tot}}+7\cos^2 i_{\mathrm{tot}})\cos(2g_1-2g_2) 
+(1-\cos i_{\mathrm{tot}})^2(1+7\cos i_{\mathrm{tot}}+7\cos^2 i_{\mathrm{tot}})\cos(2g_1+2g_2)\right] \nonumber \\
& +21e_1^2(1-\cos^2 i_{\mathrm{tot}})
\left[(1+\cos i_{\mathrm{tot}})^2\cos(4g_1-2g_2)+(1-\cos i_{\mathrm{tot}})^2\cos(4g_1+2g_2)\right] \Bigg] \Bigg\} \nonumber \\
& -\frac{1}{2\pi}\left( \frac{k_2(m_1+m_2)}{a_1^3} \right)^{1/2}
\left( \frac{45\pi}{2048} \right)
\frac{m_3}{m_1+m_2}
\left( \frac{a_1}{a_2} \right)^5
\left(1-3\frac{m_1m_2}{(m_1+m_2)^2}\right) \nonumber \\
& \times \frac{\cos i_1}{\sqrt{1-e_1^2}(1-e_2^2)^{7/2}} \frac{\sqrt{1-\cos^2 i_{\mathrm{tot}}}}{\sin i_1} \nonumber \\
& \times \Bigg\{ 2(2+3e_2^2)\cos i_{\mathrm{tot}}\left[(8+40e_1^2+15e_1^4)(3-7\cos^2 i_{\mathrm{tot}}) 
-28e_1^2(2+e_1^2)(4-7\cos^2 i_{\mathrm{tot}})\cos(2g_1) 
+147e_1^4(1-\cos^2 i_{\mathrm{tot}})\cos(4g_1)\right] \nonumber \\
& -4e_2^2\cos i_{\mathrm{tot}}(4-7\cos^2 i_{\mathrm{tot}})(8+40e_1^2+15e_1^4)\cos(2g_2) \nonumber \\
& +7e_2^2e_1^2 \Bigg[ 2(2+e_1^2)\left[(1+\cos i_{\mathrm{tot}})(5+7\cos i_{\mathrm{tot}}-28\cos^2 i_{\mathrm{tot}})\cos(2g_1-2g_2) 
-(1-\cos i_{\mathrm{tot}})(5-7\cos i_{\mathrm{tot}}-28\cos^2 i_{\mathrm{tot}})\cos(2g_1+2g_2)\right] \nonumber \\
& -21e_1^2\left[(1-2\cos i_{\mathrm{tot}})(1+\cos i_{\mathrm{tot}})^2\cos(4g_1-2g_2) 
-(1+2\cos i_{\mathrm{tot}})(1-\cos i_{\mathrm{tot}})^2\cos(4g_1+2g_2)\right] \Bigg] \Bigg\}
\end{align}
\end{tightdisplay}

\begin{tightdisplay}
\begin{align}
\left.\frac{dg_2}{dt}\right|_{\text{Hex}} ={}&
\frac{1}{2\pi} \left( \frac{k_2(m_1+m_2)}{a_1^3} \right)^{1/2}
\left( \frac{45\pi}{4096} \right)
\frac{m_1m_2}{(m_1+m_2)^2}
\left(1-3\frac{m_1m_2}{(m_1+m_2)^2}\right)
\sqrt{1+\frac{m_3}{m_1+m_2}} 
\left( \frac{a_1}{a_2} \right)^{11/2} (1-e_2^2)^{-4} \nonumber \\
&\times \Bigg\{ (4+3e_2^2)
\left[(8+40e_1^2+15e_1^4)(3-30\cos^2 i_{\mathrm{tot}}+35\cos^4 i_{\mathrm{tot}})
-140e_1^2(2+e_1^2)(1-\cos^2 i_{\mathrm{tot}})(1-7\cos^2 i_{\mathrm{tot}})\cos(2g_1) \right.\nonumber \\
&\left. +735e_1^4(1-\cos^2 i_{\mathrm{tot}})^2\cos(4g_1)\right] 
-(2+5e_2^2) \Bigg[2(8+40e_1^2+15e_1^4)(1-\cos^2 i_{\mathrm{tot}})(1-7\cos^2 i_{\mathrm{tot}})\cos(2g_2) \nonumber \\
& -28e_1^2(2+e_1^2)
\left[(1+\cos i_{\mathrm{tot}})^2(1-7\cos i_{\mathrm{tot}}+7\cos^2 i_{\mathrm{tot}})\cos(2g_1-2g_2)
+(1-\cos i_{\mathrm{tot}})^2(1+7\cos i_{\mathrm{tot}}+7\cos^2 i_{\mathrm{tot}})\cos(2g_1+2g_2)\right] \nonumber \\
& -147e_1^4(1-\cos^2 i_{\mathrm{tot}})
\left[(1+\cos i_{\mathrm{tot}})^2\cos(4g_1-2g_2)+(1-\cos i_{\mathrm{tot}})^2\cos(4g_1+2g_2)\right] \Bigg] \Bigg\} \nonumber \\
& - \frac{1}{2\pi} \left( \frac{k_2(m_1+m_2)}{a_1^3} \right)^{1/2}
\left( \frac{45\pi}{2048} \right)
\frac{m_3}{m_1+m_2}
\left( \frac{a_1}{a_2} \right)^5
\left(1-3\frac{m_1m_2}{(m_1+m_2)^2}\right)
\cos i_2 (1-e_1^2)^{-1/2}(1-e_2^2)^{-7/2} \frac{\sqrt{1-\cos^2 i_{\mathrm{tot}}}}{\sin i_1} \nonumber \\
&\times \Bigg\{
2(2+3e_2^2)\cos i_{\mathrm{tot}} 
\left[(8+40e_1^2+15e_1^4)(3-7\cos^2 i_{\mathrm{tot}})
-28e_1^2(2+e_1^2)(4-7\cos^2 i_{\mathrm{tot}})\cos(2g_1)
+147e_1^4(1-\cos^2 i_{\mathrm{tot}})\cos(4g_1)\right] \nonumber \\
& -4e_2^2\cos i_{\mathrm{tot}}(4-7\cos^2 i_{\mathrm{tot}})(8+40e_1^2+15e_1^4)\cos(2g_2) \nonumber \\
& +7e_2^2e_1^2 \Bigg[ 2(2+e_1^2)
\left[(1+\cos i_{\mathrm{tot}})(5+7\cos i_{\mathrm{tot}}-28\cos^2 i_{\mathrm{tot}})\cos(2g_1-2g_2)
-(1-\cos i_{\mathrm{tot}})(5-7\cos i_{\mathrm{tot}}-28\cos^2 i_{\mathrm{tot}})\cos(2g_1+2g_2)\right] \nonumber \\
& -21e_1^2 \left[(1-2\cos i_{\mathrm{tot}})(1+\cos i_{\mathrm{tot}})^2\cos(4g_1-2g_2)
-(1+2\cos i_{\mathrm{tot}})(1-\cos i_{\mathrm{tot}})^2\cos(4g_1+2g_2)\right] \Bigg] \Bigg\}
\end{align}
\end{tightdisplay}

\noindent \textbf{Angular momentum evolution:}
\noindent
\begin{tightdisplay}
\begin{align}
\left.\frac{dG_1}{dt}\right|_{\text{Hex}} ={}&
\frac{315}{8192} \frac{a_1^4 k_2 m_1 m_2 m_3}{a_2^5 (m_1+m_2)} \left(1-3\frac{m_1m_2}{(m_1+m_2)^2}\right)
\frac{e_1^2}{(1-e_2^2)^{7/2}} \times \Bigg\{ \nonumber \\
& 4e_2^2\sin(2g_2)\left[(2+e_1^2)\cos(2g_1)\left(\cos i_{\mathrm{tot}} + 7(-3\cos i_{\mathrm{tot}} + 4\cos^3 i_{\mathrm{tot}})\right)
+ 84e_1^2\cos(4g_1)\cos i_{\mathrm{tot}}(1-\cos^2 i_{\mathrm{tot}})\right] \nonumber \\
& + \sin(2g_1)\Big[
-84e_1^2\cos(2g_1)(1-\cos^2 i_{\mathrm{tot}})
\left(2e_2^2\cos(2g_2)(3+2\cos^2 i_{\mathrm{tot}}-1) + 2(2+3e_2^2)(1-\cos^2 i_{\mathrm{tot}})\right) \nonumber \\
& + (2+e_1^2)\Big(
-2e_2^2\cos(2g_2)\left(5+8\cos^2 i_{\mathrm{tot}}-4 + 7(1 - 8\cos^2 i_{\mathrm{tot}} + 8\cos^4 i_{\mathrm{tot}})\right) \nonumber \\
& -4(2+3e_2^2)(5+14\cos^2 i_{\mathrm{tot}}-7)(1-\cos^2 i_{\mathrm{tot}})
\Big)
\Big]
\Bigg\}
\end{align}
\end{tightdisplay}
\begin{tightdisplay}
\begin{align}
\left.\frac{dG_2}{dt}\right|_{\text{Hex}} ={}&
-\frac{45}{4096} \frac{a_1^4 k_2 m_1 m_2 m_3}{a_2^5 (m_1+m_2)^3} \frac{e_2^2}{(1-e_2^2)^{7/2}} (m_1^2 - m_1m_2 + m_2^2) \times \Bigg\{ \nonumber \\
& \sin(2g_2)\Big[
7e_1^2(2+e_1^2)\cos(2g_1)\left(5 + 8\cos^2 i_{\mathrm{tot}} - 4 + 7(1 - 8\cos^2 i_{\mathrm{tot}} + 8\cos^4 i_{\mathrm{tot}})\right) \nonumber \\
& + (1-\cos^2 i_{\mathrm{tot}})\left(
147e_1^4\cos(4g_1)(3 + 2\cos^2 i_{\mathrm{tot}} -1)
+ (8+40e_1^2+15e_1^4)(5 + 14\cos^2 i_{\mathrm{tot}} - 7)
\right)
\Big] \nonumber \\
& + 14e_1^2\cos(2g_2)\Big[
-(2+e_1^2)\left(\cos i_{\mathrm{tot}} + 7(-3\cos i_{\mathrm{tot}} + 4\cos^3 i_{\mathrm{tot}})\right)\sin(2g_1)
-42e_1^2\cos i_{\mathrm{tot}}(1-\cos^2 i_{\mathrm{tot}})\sin(4g_1)
\Big]
\Bigg\}
\end{align}
\end{tightdisplay}
\noindent
\textbf{Longitude of ascending node evolution:}

\begin{tightdisplay}
\begin{align}
\left.\frac{dh_1}{dt}\right|_{\text{Hex}} ={}&
\frac{1}{2\pi} \left( \frac{k_2(m_1+m_2)}{a_1^3} \right)^{1/2}
\left( -\frac{45\pi}{2048} \right)
\frac{m_3}{m_1+m_2}
\left( \frac{a_1}{a_2} \right)^5
\left( 1-3\frac{m_1m_2}{(m_1+m_2)^2} \right)
\frac{1}{(1-e_{1}^{2})^{1/2}(1-e_{2}^{2})^{7/2}} \frac{\sin(i_{\text{tot}})}{\sin(i_{1})} \times \Bigg[ \nonumber \\
& 2(2+3e_{2})\cos(i_{\text{tot}})\Big[(8+40e_{1}^{2}+15e_{1}^{4})(3-7\cos^{2}(i_{\text{tot}})) \nonumber \\
& -28e_{1}^{2}(2+e_{1}^{2})(4-7\cos^{2}(i_{\text{tot}}))\cos(2g_{1}) \nonumber \\
& +147e_{1}^{4}\sin^{2}(i_{\text{tot}})\cos(4g_{1})\Big] \nonumber \\
& -4e_{2}^{2}\cos(i_{\text{tot}})(4-7\cos^{2}(i_{\text{tot}}))(8+40e_{1}^{2}+15e_{1}^{4})\cos(2g_{2}) \nonumber \\
& +7e_{1}^{2}e_{2}^{2}\Big[ 2(2+e_{1}^{2})\Big[(1+\cos(i_{\text{tot}}))(5+7\cos(i_{\text{tot}})-28\cos^{2}(i_{\text{tot}}))\cos(2g_{1}-2g_{2}) \nonumber \\
& -(1-\cos(i_{\text{tot}}))(5-7\cos(i_{\text{tot}})-28\cos^{2}(i_{\text{tot}}))\cos(2g_{1}+2g_{2})\Big] \nonumber \\
& -21e_{1}^{2}\Big[(1-2\cos(i_{\text{tot}}))(1+\cos(i_{\text{tot}}))^{2}\cos(4g_{1}-2g_{2}) \nonumber \\
& -(1+2\cos(i_{\text{tot}}))(1-\cos(i_{\text{tot}}))^{2}\cos(4g_{1}+2g_{2})\Big] \Big] \Bigg]
\end{align}
\end{tightdisplay}
\begin{figure*}
 \begin{center}
 \includegraphics[width=\textwidth]{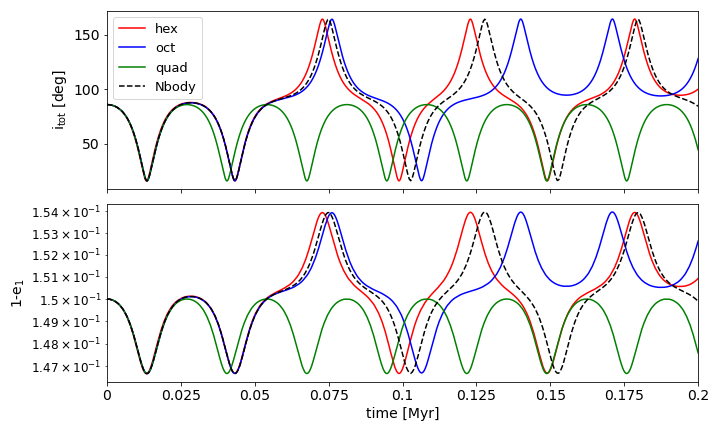}
 \caption{A comparison for the quadrupole, octupole, and hexadecapole orders with direct Nbody integration performed with {\tt REBOUND} \citep{rebound, reboundias15} for an example system. The initial conditions were $m_1=0.80$~M$_{\odot}$, $m_2=0.20$~M$_{\odot}$, $m_3=3.0\times10^{-4}$~M$_{\odot}$, $a_1=0.8$~au, $a_2=8.5$~au, $\omega_1=0^{\circ}$, $\omega_2=100^{\circ}$, $e_1=0.85$, $e_2=0.08$, and $i_{\rm tot}=85.85^{\circ}$. For more examples, \citet{Will+17} highlights several systems showcasing the difference between the octupole and hexadecapole.}
 \label{fig:expansion}
 \end{center}
\end{figure*}

%\clearpage
%\section{Hexadecapole term in Hamiltonian}\label{appendix:hexham} 

%\input{VHS_Appendix_Ham.tex}
\clearpage
\section{Tertiary Spin Equations}\label{appendix:spin}
%\providecommand{\standalone}{}
%\ifdefined\standalone
%\documentclass[12pt]{article}
%\usepackage{amsmath}
%\usepackage{graphicx}
%\usepackage[margin=1in]{geometry}
%\begin{document}
%\fi

%\begin{center}
%\Large \textbf{Appendix B: Tertiary Spin}
%\end{center}
%\vspace{1em}
%\small
%\noindent
%Along with EKL, the evolution of hierarchical triples is also governed by general relativity. The first post-Newtonian order (1PN) drives the precession of the inner and outer orbits, inducing resonant-like behavior in the chaotic evolution of the eccentricity when the octupole timescale and the 1PN timescale are similar. 
%Because EKL can lead to very small distances of periapsis, it becomes prudent to consider the effects of tides and rotation. If the tidal precession timescale is similar to the EKL timescale, eccentricity excitations are suppressed, and the inner binary can shrink and circularize into a stable configuration decoupled from the tertiary. If the EKL timescale is faster than the tides, the octupole can create an almost radial orbit, and the objects merge before the tidal effects have time to stabilize the system. We display the equations governing the spin evolution of the tertiary companion below.\\
\noindent
We define the spin of the tertiary body as follows:
\renewcommand{\theequation}{D\arabic{equation}}
\setcounter{equation}{0}

\begin{equation}
    \vec{S}_{3} = (s_{3e}, s_{3q}, s_{3h}) \ .
\end{equation}
We define the variable $n_{3}$ as follows:
\begin{equation}
    n_{3}=\frac{2\pi}{P_{2}} \ , 
\end{equation}
where $P_{2}$ is the period of the outer orbit. We define the reduced mass $\mu_{3}$ as
\begin{equation}
    \mu_{3}=\frac{m_{3}(m_{1}+m_{2})}{m_{1}+m_{2}+m_{3}} \ .
\end{equation}
We then define the moment of inertia $I_{n3}$ as 
\begin{equation}
    I_{n3}=\alpha_{3}m_{3}R_{3}^{2} \ .
\end{equation}
Next, we define the tidal dissipation rate as
\begin{equation}
    tF_{3} = TV_{3}\Big(\frac{a_{2}}{R_{3}}\Big)^{8}\frac{m_{3}^{2}}{9(m_{1}+m_{2}+m_{3})(m_{1}+m_{2})(1+2kL_{3})^{2}} \ ,
\end{equation}
where $TV_{3}$ is the viscous timescale for the tertiary body and $kL_{3}$ is defined as 
\begin{equation}
    kL_{3}=\frac{1}{2}\frac{Q_{3}}{1-Q_{3}} \ ,
\end{equation}
where $Q_{3}$ is from \citet{Eggleton98}. We define the components of the spin $V_{3}, W_{3}, X_{3}, Y_{3},$ and $Z_{3}$ below. 
\begin{equation}
    V_{3}=9\bigg(\Big(1+\frac{15}{4}e_{2}^{2}+\frac{15}{8}e_{2}^{4}+\frac{5}{64}e_{2}^{6}\Big)\Big(\frac{1}{1-e_{2}^{2}}\Big)^{\frac{13}{2}}-\frac{11s_{3h}}{18n_{3}}\Big(1+\frac{3}{2}e_{2}^{2}+\frac{1}{8}e_{2}^{4}\Big)\Big(\frac{1}{1-e_{2}}\Big)^{5}\bigg)\frac{1}{tF_{3}} \ ,
\end{equation}
\begin{equation}
    W_{3}=\bigg(\big(1+\frac{15}{2}e_{2}^{2}+\frac{45}{8}e_{2}^{4}+\frac{5}{16}e_{2}^{6}\Big)\Big(\frac{1}{1-e_{2}^{2}}\Big)^{\frac{13}{2}}-\frac{s_{3h}}{n_{3}}\Big(1+3e_{2}^{2}+\frac{3}{8}e_{2}^{4}\Big)\Big(\frac{1}{1-e_{2}^{2}}\Big)^{5}\bigg)\frac{1}{tF_{3}} \ ,
\end{equation}
\begin{equation}
    X_{3}=-\Big(\frac{R_{3}}{a_{2}}\Big)^{5}kL_{3}\frac{m_{1}+m_{2}}{\mu_{3}n_{3}}\frac{s_{3h}s_{3e}}{(1-e_{2}^{2})^{2}}-\frac{s_{3q}}{2n_{3}tF_{3}}\Big(1+\frac{9}{2}e_{2}^{2}+\frac{5}{8}e_{2}^{4}\Big)\Big(\frac{1}{1-e_{2}^{2}}\Big)^{5} \ ,
\end{equation}
\begin{equation}
    Y_{3}=-\Big(\frac{R_{3}}{a_{2}}\Big)^{5}kL_{3}\frac{m_{1}+m_{2}}{\mu_{3}n_{3}}\frac{s_{3h}s_{3q}}{(1-e_{2}^{2})^{2}}+\frac{s_{3e}}{2n_{3}tF_{3}}\Big(1+\frac{3}{2}e_{2}^{2}+\frac{1}{8}e_{2}^{4}\Big)\Big(\frac{1}{1-e_{2}^{2}}\Big)^{5} \ ,
\end{equation}
\begin{equation}
    Z_{3}=\Big(\frac{R_{3}}{a_{2}}\Big)^{5}kL_{3}\frac{m_{1}+m_{2}}{\mu_{3}n_{3}}\bigg(\frac{2s_{3h}^{2}-2s_{3q}^{2}}{2(1-e_{2}^{2}}+\frac{15k^{2}(m_{1}+m_{2})}{a_{2}^{3}}\Big(1+\frac{3}{2}e_{2}^{2}+\frac{1}{8}e_{2}^{4}\Big)\Big(\frac{1}{1-e_{2}^{2}}\Big)^{5}\bigg) \ .
\end{equation}
%\ifdefined\standalone
%\end{document}
%\fi

%\clearpage
\section{Timestamp figure for the case where $m_{3}\approx 12M_{J}$}\label{appendix:m312}
\begin{figure*}[t]
 \centering
 \includegraphics[width=\textwidth]{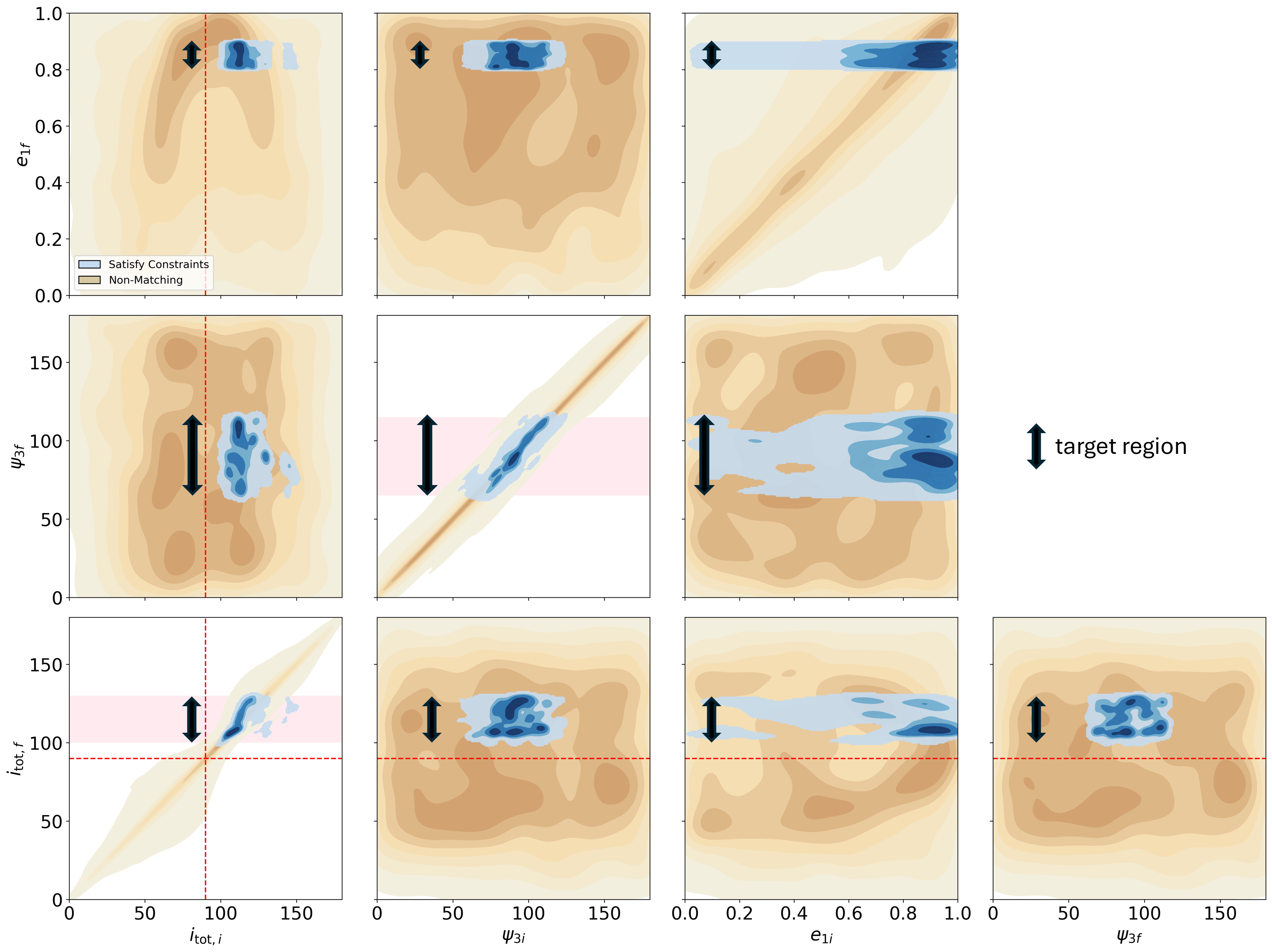}
 \caption{Time evolution of 2000 sample systems with $q=\frac{m_{\rm VHSA}}{m_{\rm VHSB}}\in[0.9,1.0]$ and $m_{3}\in[11$M$_{\rm J},13$~M$_{\rm J}]$ corresponding to a deuterium inert-model for VHS 1256b. The red dashed lines indicate $90^{\circ}$. The pink shaded boxes highlight the "target region", where we define the target region based on the observational constraints, $e_{1}\in[0.8,0.9]$, $i_{\rm tot}\in[100^{\circ},130^{\circ}]$, and $\psi_{3}\in[65^{\circ},115^{\circ}]$. For each system, $400$ timestamps are considered ranging from $120-160$~Myr. If at a given timestamp, the system obeys all three observational constraints, the point turns blue. If it does not, it turns beige. Once all timestamps are considered for each system, the blue points are accumulated and displayed by the blue density clouds, while the beige points are accumulated and displayed by the beige density clouds.}
 \label{fig:timestamps12}
\end{figure*}

\clearpage
\bibliography{VHS}{}
\bibliographystyle{aasjournal}
\end{document}